\documentclass[twocolumn]{aastex631}
\usepackage{array}
\usepackage{multirow}
\newcolumntype{P}[1]{>{\centering\arraybackslash}p{#1}}
\usepackage{graphicx}	
\usepackage{amsmath}	
\usepackage{amssymb}	
\usepackage{comment}
\usepackage{newtxtext,newtxmath}
\usepackage{booktabs}
\usepackage[graphicx]{realboxes}
\usepackage{rotating}
\usepackage{nicefrac}
\usepackage[maxfloats=256]{morefloats}
\usepackage{makecell}
\newcommand{\angstrom}{\mbox{\normalfont\AA}}

\usepackage{amsmath} 

\received{July 9, 2025}
\revised{October 22, 2025}
\accepted{November 5, 2025}

\shorttitle{Continuum Reverberation Mapping of 18 AGN time lags}
\shortauthors{Miller et al.}
\begin{document}

\title{Continuum Reverberation Mapping of 18 AGN Over Four Years}
\author[0000-0001-8475-8027]{Jake A. Miller}
\affiliation{Texas A\&M University, Department of Physics \& Astronomy, 400 Bizzell St, College Station, TX 77845, USA}
\affil{Wayne State University, Department of Physics \& Astronomy, 666 W Hancock St, Detroit, MI 48201, USA}
\author[0000-0002-8294-9281]{Edward M. Cackett}
\affiliation{Wayne State University, Department of Physics \& Astronomy, 666 W Hancock St, Detroit, MI 48201, USA}

\author[0000-0002-2816-5398]{Misty C.\ Bentz}
\affiliation{Department of Physics and Astronomy,
Georgia State University,
Atlanta, GA 30303, USA}
\author[0000-0002-2908-7360]{Michael R. Goad}
\affiliation{School of Physics and Astronomy, University of Leicester, University Road, Leicester, LE1 7RH, UK}
\author[0000-0003-0944-1008]{Kirk T. Korista}
\affiliation{Western Michigan University,
Department of Physics,
Kalamazoo, MI, 49008-5252 USA}
\author{Ian M. McHardy}
\affiliation{School of Physics and Astronomy, University of Southampton, Highfield, Southampton SO17 1BJ, UK}


\begin{abstract}
Continuum reverberation mapping probes the size scale of the optical continuum-emitting region in active galactic nuclei (AGN). The source of this emission has long been thought to originate from the accretion disk, but recent studies suggest the broad line region (BLR) may significantly contribute to both the observed flux and continuum interband delays. We monitored 18 AGN over four years of observations to acquire high quality optical continuum light curves, measuring time lags between different photometric band and determining continuum emission sizes for each AGN. 
We add this sample to existing lag measurements to test the correlation between continuum lags at $5100\Angstrom$ ($\tau_{5100}$) and $5100\Angstrom$ luminosity ($L_{5100}$). We observe that $\tau_{5100}\propto L_{5100}^{0.4}$, broadly consistent with the theoretical expectations of $\tau \propto L^{1/2}$ expected for continuum reverberation from either the accretion disk or the BLR.
\end{abstract}

\section{Introduction}
Reverberation mapping (RM) has been used to study the inner regions of Active Galactic Nuclei (AGN) for decades \citep{blandford1982}. This technique probes the structure of the AGN system indirectly using light echoes. One of the most intriguing {applications} of RM is {for measuring the radial temperature profile of} the accretion disk. In this regime, the simplest model (also known as the lamppost model) {assumes} that {variable and highly energetic X-rays are produced by {the electron corona of a} supermassive black hole (SMBH) and emitted isotropically, illuminating and depositing energy into the thermal accretion disk (e.g., \cite{cackett2007})}. The disk then reprocesses these photons, absorbing some energy and re-emitting photons at longer wavelengths dependent on the temperature of the disk. {This predicts a time lag for ultraviolet and optical continuum emission that is dependent on the light travel time between the X-ray source and the disk.} {We refer to this specific reprocessing scenario as continuum RM.} For a review of RM, see \cite{cackett2021_review}.

This variable optical continuum emission has been observed in many AGN, and {in general,} measured time delays between different {continuum bands} follow the wavelength scaling relationship expected for an optically-thick geometrically-thin accretion disk \citep{1973A&A....24..337S} (hereafter referred to as `thin disk'). {The thin disk model predicts that the time lags $\tau$ between observed wavelengths $\lambda$ should be related as $\tau \propto \lambda^{4/3}$ \citep{collier1999}}. However, there are several important predictions that many AGN do not adhere to. If the X-rays are driving this variability {at longer wavelengths}, we would expect to see a strong relation between the observed X-rays and the reprocessed accretion disk emission, but in general the relationship is far weaker than expected {and sometimes nonexistent} on both short and long timescales \citep{breedt2009, shappee2014, edelson2017, edelson2019, fausnaugh2016, kara2021}.

{There are potential explanations for the lack of correlation. X-ray reprocessing of an inhomogeneous disk may disconnect the X-ray from the UV/Optical \citep{ren2024}. A low-to-moderate X-ray correlation might even be expected in the case of a variable corona height \citep{panagiotou2022}. More complex coronal geometries involving a far-UV reprocessor may also complicate correlations \citep{gardner2017}. However, magnetohydrodynamic simulations predict that the X-ray light curves are primarily uncorrelated with the UV/Optical and do not drive the variability seen in these wavelengths \citep{secunda2024}. Currently, there is no broadly accepted explanation for the low X-ray to UV/optical correlation observed in most AGN, and even further division as to whether this can be reconciled with the expectations of the lamppost model. }

In addition, the length of the lags between {the continuum} bands is larger than anticipated, implying accretion disks are $\sim$3 times greater in size than predicted {by simple scaling relations} \citep{fausnaugh2016, edelson2019, cackett2020, kara2021, guo2022, Miller2023}. {More sophisticated disk-dominated models including effects such as General Relativity and the height of the corona can fit the lags well \citep{kammoun21b, kammoun2023}, although they neglect contributions from additional reprocessing sites that are known to be present}. While {inferred disk sizes for the majority of AGN seem to disagree with thin disk predictions, there exist some AGN that do agree with these predictions \citep{homayouni2019, jha2022} .} 
{While some evidence exists linking agreement with thin disk theory to AGN luminosity \citep{Li2021}, it is unknown if any other parameters can influence this relationship.}

{Alternative explanations for AGN variability on both shorter timescales ($\sim$days) and longer timescales ($\sim$years) exist. For example, the corona-heated accretion-disk reprocessing ( or CHAR) model associates the temperature fluctuations of the disk with magnetic coupling from the X-ray corona \citep{sun2020a}. Tests with simulated data modeled off the Stripe-82 region from the Sloan Digital Sky Survey are able to recover the input SMBH mass and bolometric luminosity, as well as predict accurate continuum reprocessing sizes \citep{sun2020b}. In a similar vein, \cite{neustadt2024} invoke coherent temperature fluctuations inherent to the disk as responsible for the variations. Applying other corrections, such as color correction, implementing the effect of disk winds, and disk truncation at radii larger than the innermost stable circular orbit can also correct theoretical sizes to match observed disk sizes \citep{zdziarski2022}. }

{The lamppost model may not be sufficient to explain all observed AGN continuum variability. However, it is the most widely used and understood model for continuum {RM}, and while not sufficient for all AGN there are some for which it can explain the observed lags \citep{ jha2022, homayouni2019}. In addition, it is possible that there is not a single solution for all observed discrepancies. The different environments of each AGN likely play a role in the observed lags. As such, we continue to {utilize this simple} model and draw attention to its shortcomings when applicable.}

Previous studies typically either are extremely detailed {multi}-wavelength looks into a single or handful of AGN \citep{derosa2015, kara2021, edelson2019, cackett2018, cackett2020} or {involve studies based on large quasar samples} taken from all-sky surveys such as the Zwicky Transient Facility (ZTF) and Pan-STARRS \citep{guo2022, 2022MNRAS.511.3005J, 2017ApJ...836..186J}. The former is the best way to completely understand the AGN system, but {is} difficult and time-consuming to organize and perform, and as such only a {few} AGN have been studied in this capacity. The latter method has great potential to significantly add to the number of AGN {that are studied}, but most surveys currently lack observation cadence and/or wavelength coverage to probe the entire disk, leading to {large uncertainties in individual results. However, they probe a broader range of AGN properties than the nearby Seyfert sample.} High cadence observations with full optical coverage are ideal for analyzing the outer accretion disk, and with the advent of remote robotic observatories, {observational campaigns of this nature} have become more feasible. 

\begin{figure}
\centering
\includegraphics[trim=16cm 0cm 16cm 0cm, clip = true, scale = 0.3]{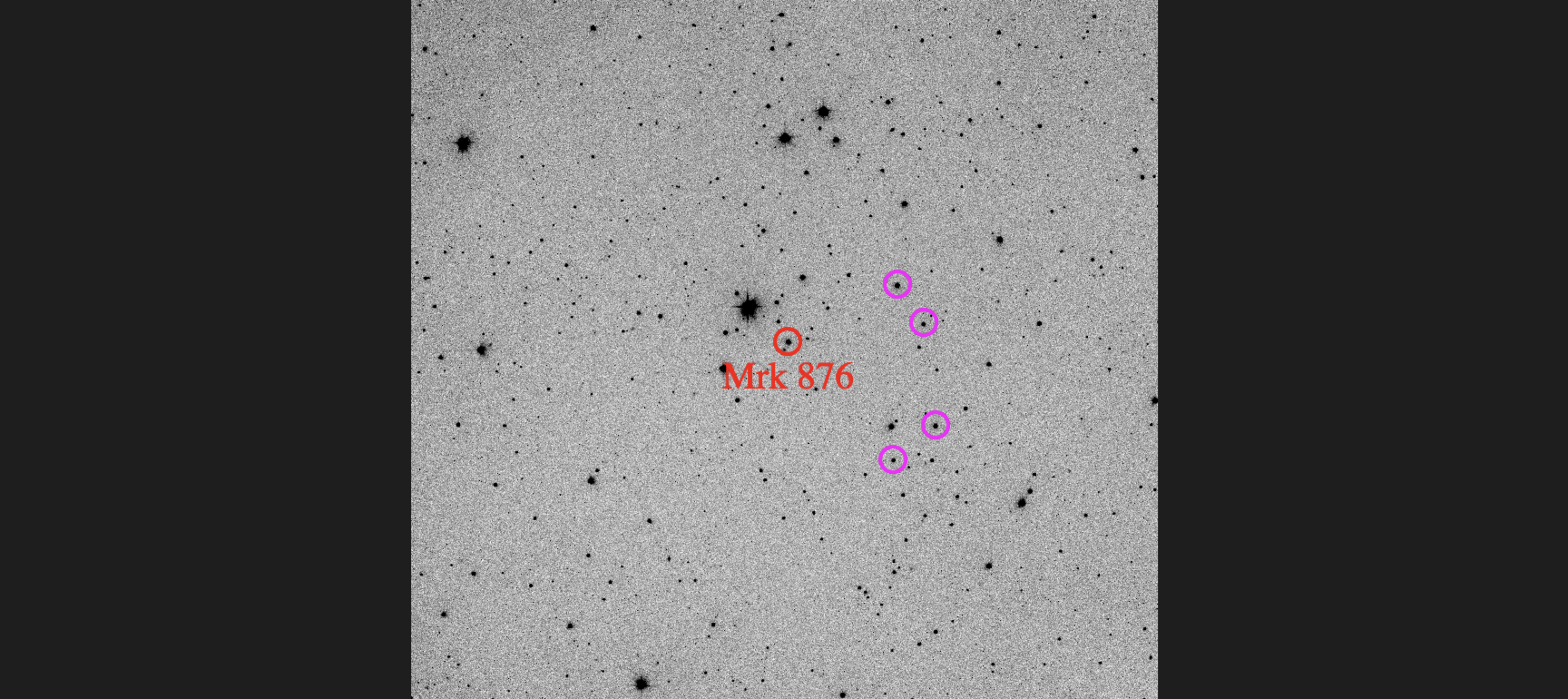}
\caption{{Example $g$-band observation of Mrk~876 taken with Zowada. {The red circle indicates the AGN's location, and the magenta circles are the four~comparison stars used for the $griz$ relative photometry. The plotted stars had the lowest fractional standard deviation out of 25~stars tested. See Sec.~\ref{sec:methods} for details.}}}
\label{fig:ZW_example}
\end{figure}

This is the case for the Dan Zowada Memorial Observatory (Zowada), a 0.5-m robotic observatory {that at the time of these observations was located outside of Rodeo, New Mexico. Zowada is equipped with the Sloan $ugri$ and Pan-STARRS $z_s$ filters (hereafter referred to as $z$) and has a field of view of $30\times30$~arcminutes. With a pixel scale of 0.89 pixel/arcsecond, it is capable of performing the intensive monitoring required to recover excellent time lags. {An example image of a $g$-band observation of Mrk~876 is presented in Fig. \ref{fig:ZW_example}.} For an in-depth overview of Zowada characteristics, please see \cite{carr2022}.} 

{Zowada} has {monitored} more than 20~AGN over the past several years. Zowada and other remote observatories occupy a niche between the focused single-target studies and all-sky surveys. These observatories can provide a significant sample of AGN with high cadence observations across the entire optical spectrum, {necessary} to better understand the connection between observable parameters of the AGN system and the measured time lags. 

One application of these studies is to determine the {primary} source of optical variability in AGN. While previously it was expected that the accretion disk provides {the bulk of} this variability, recent studies have proposed {that optical AGN variability has significant contribution from the broad line region (BLR)} \citep{netzer2022, montano2022, guo2022}. Previous studies have found lengthened {lags around $3500\Angstrom$,} likely originating from the Balmer {continuum} feature in the BLR \citep{fausnaugh2016, cackett2018, edelson2019, jiang2024}. This {lag excess} has also been {reported} in frequency-resolved analysis techniques \citep{cackett22, Lewin2023} which have been shown to need an additional, larger reprocessor consistent with the BLR. {The flattening of the optical spectrum cannot easily be explained with only the disk's emission, but is easily explained with the inclusion of the diffuse continuum originating from the BLR \citep{korista2019}.} 

Finally, studies have shown that {continuum emission originating from the BLR could lengthen measured accretion disk lags \citep{lawther2018, korista2001, netzer2022}. BLR continuum} emission has often been invoked to explain larger than expected disk sizes. {However, in some studies this effect is not seen \citep{sharp2024}, implying that other phenomena may be responsible.} {While the lack of X-ray/UV observations does not allow us to constrain the true shape of the lag profile for these objects, we can still measure the lag amplitude and thus get a bearing on the size scale of the primary reprocessing region(s).} 

{If variable reprocessed continuum emission from the BLR dominates the interband continuum delays,} then it is expected that the relationship between the accretion disk size and luminosity at 5100$\mathrm{\mathring{A}}$ ($L_{5100}$) should mimic the well established relationship between BLR radius ($\mathrm{R_{BLR}}$) and $L_{5100}$ \citep{bentz2013}{, where $\mathrm{R_{BLR}} \propto L_{5100}^{1/2}$.} 
{In the continuum RM model, the lags are predicted to scale to with SMBH mass $(M)$ and mass accretion rate $(\dot{M})$, following $\tau_0 \propto (M\dot{M})^{1/3}$. Previous work \citep{montano2022, guo2022} has made the assumption of a constant bolometric correction factor (i.e., $L_{5100} \propto \dot{M}$) and found an empirical relationship between $M$ and $L_{5100}$ in order to write this relationship solely in terms of $L_{5100}$. However, we also note that Eq.~3.12 from \cite{1973A&A....24..337S} predicts that the optical luminosity of a sub-Eddington accretion rate black hole is related to its mass and mass accretion rate $\textrm{L} \propto (\textrm{M}_{\textrm{BH}}\dot{\textrm{M}})^{2/3}$. This can be combined with Eq.~\ref{eq:lageq} to show that $\tau \propto \textrm{L}^{1/2}$ -- the same relation expected for the BLR. Even in the case of the constant bolometric luminosity assumption, the resulting scaling is nearly indistinguishable from the BLR relation. With this study, we supplement the sample examined in previous studies and expand the parameter range probed in order to investigate the dependence of continuum lags on the optical luminosity.}


The paper is structured as follows. Section \ref{sec:methods} contains descriptions of our observations and data reduction methods. Section \ref{sec:analysis} follows our initial analysis of the data, including the calculation of time lags using several different methods. Section \ref{sec:discussion} is our discussion of these results, and the implications they contain. 
{Finally, Section \ref{sec:conclusions} summarizes our results and presents our conclusions.} We adopt a standard $\Lambda$CDM cosmology with $H_0$~=~72~km/s/Mpc, $\Omega_\Lambda$~=~0.7, and $\Omega_M$~=~0.3. {For our filters, we assume central bandpasses of $3540\Angstrom, 4770\Angstrom, 6215\Angstrom, 7545\Angstrom,$ and $8700\Angstrom$ for the $ugriz$ filters respectively.}
\section{Methods}
\label{sec:methods}
{The AGN in this study} were observed with the Zowada {Observatory} over four years, from 2018-2022. The observing season for Zowada starts in late summer/early fall {after monsoon season} {each} year, when the conditions surrounding the telescope {become ideal for} operation. {Typically, this allows about 10 months of viewing, with some variation depending on the weather.} 

{The first two years of observing were less consistent in the number of images and exposure times for each observation and each filter. This limited the number {of AGN} observed in the first year (2018-2019) especially, as we report only 4~AGN with robust enough light curves to measure time lags. As we experimented with different observation configurations with Zowada, we determined the number of exposures for each filter per night of observation. For all images in all filters we use exposure times of 300~seconds. {Multiple consecutive images are taken in each filter to boost the signal-to-noise ratio and to mitigate the effect of cosmic rays that may be captured during an exposure.} Exposures in $g$, $r$, and $i$ bands required only 2~observations each. We observe the $z$-band 3~times and $u$ band 4~times. All images are dark, bias, and flat-field corrected to remove observatory contamination. Due to the limited sensitivity of the $u$-band, we also stack these images together to enhance the signal to noise ratio.}

The light curves for each AGN are obtained using differential photometry. For each AGN, we {initially select} around 25 comparison stars that are observable in {the $griz$} bands. We assume that the combined flux from these stars is constant. The AGN light curve is then calculated by observing how the total brightness changes over time, allowing us to recover the relative AGN variability. {We take the maximum fractional standard deviation (STD) of all comparison stars to be the systematic uncertainty that is added to the AGN's measured uncertainty in quadrature. To reduce this amount, we selectively remove stars with a large fractional STD from the initial 25~stars for each AGN until between 2-4~stars remain. The same comparison stars are used for the $griz$ filters.} Different comparison stars are chosen for the $u$ band, since typically stars are fainter in this band, but otherwise the procedure for this band is the same as above.

The stars and AGN are found for each image using the photutils \citep{photutils2021} module DAOStarFinder. A circular aperture and {background} annulus are created around the AGN and the chosen comparison stars. The aperture {radius} is 5~pixels, corresponding to $\sim$4.5~arcseconds. The annuli have an inner radius of 20~pixels and an outer radius of 30~pixels. The median background is measured within the annulus and is scaled to the area of the {source} aperture and subtracted. {Observations for each night are averaged unless more than 3~hours have passed between observations, in which case it becomes a separate data point.} {The light curve for each band is normalized so that the AGN has a mean count rate of~1}. Once assembled, we perform sigma-clipping on the light curves to remove errant points due to weather conditions or stray cosmic rays. This process involves {removing} observations that occurred {on a night when any comparison} star deviates from its mean by more than {5}$\sigma$. In addition, each observation is compared against a moving boxcar average multiplied by a {scale factor}, typically between 5 and 20, {to remove large jumps between consecutive nights of observations that are atypical of AGN variability patterns and likely originate due to poor observing conditions. The exact boxcar width and height was scaled manually depending on average cadence and quality of the light curve. Data points with 0.1 relative counts difference between consecutive nights of observation were also removed due to exhibiting variability on timescales inconsistent with AGN. This difference was adjusted to 0.2 counts for the $u$ band and occasionally less for redder bands with lower variability amplitude.} The {final} $g$-band light curves are {presented} in Fig.~\ref{fig:LC_1}. {See the Appendix for the full collection of light curves, as well as Tables \ref{table:year1_lt}, \ref{table:year2_lt}, \ref{table:year3_lt}, and \ref{table:year4_lt} for detailed light curve information on all of the objects in the 2018-2019, 2019-2020, 2020-2021, and 2021-2022 observing seasons.}

Once light curves {have been} constructed, we {analyzed} each AGN to determine if it {was} observed for a minimum duration and at a high enough cadence to recover accurate time lags. We also {calculated} the fractional variability ($F_{var}$) of each band to determine if the AGN {is} variable during that particular observing season {\citep{edelson1990,rodriguez-pascual1997, vaughan2003}.} We {applied} the following criteria and remove AGN from our sample that do not meet these {criteria}:
\begin{itemize}
  \item Average gap between observations {is} less than three days {\citep[][]{fausnaugh2016, cackett2021_review}}.
  \item {The AGN is observed} for more than three times the length of the expected lag signal, {corresponding to a 90~day observation duration} {\citep{horne2004}}.
  \item At least two bands {are} variable enough to measure {robust} lags with {$F_{var}$ $\geq$ 0.03}. 
\end{itemize}
{These criteria result in 18 unique AGN for our final sample, nine of which we observe for two separate observing seasons and one (Mrk~817) for three seasons. A total of 29 $ugriz$ light curves are collected.}
In addition, many objects {were} observed as part of coordinated observation campaigns. We indicate these objects here. {In the 2018-2019 observation season, we observed Mrk~110 \citep{vincentelli2022}, Mrk~142 \citep{cackett2020}, and Mrk~817 \citep{kara2021}. In the 2019-2020 observation season, we observed Mrk~335 \citep{kara2023}. In the 2020-2021 observation season, we observed MCG+8-11-11 (Kynoch et al., in prep.) and Mrk~817 (Montano et al., in prep.). In the 2021-2022 observation season, we observed Mrk~817 (Montano et al., in prep.).} 

{We observed several objects that we cut from our final sample due to a lack of variability. AGN variability patterns shift and adjust over time and an element of luck is involved when observing, especially with less powerful observatories like Zowada \citep{horne2004, cackett2021_review}. The four AGN we remove due to this criteria are Mrk~876 (observed 2018-2019), PG0953+414 (2019-2020), PG1229+204 (2019-2020) and 1E07546+3928 (2020-2021). The final list of objects we use in this study and the observing seasons they were recorded in are listed in Table~\ref{table:quicklook_lt}.}

\begin{deluxetable}{cc}
\tablecolumns{2}
\tablecaption{Zowada Survey Targets}
\label{table:quicklook_lt}
\tablehead{
Object & Seasons Observed}
\startdata
3C390.3 & \makecell{2019-2020 \\2020-2021 }  \\
\hline
Arp 151 & \makecell{2019-2020 }  \\
\hline
MCG+8-11-11 & \makecell{2020-2021 \\2021-2022 }  \\
\hline
Mrk 6 & \makecell{2019-2020 \\2020-2021 }  \\
\hline
Mrk 50 & \makecell{2020-2021 \\2021-2022 }  \\
\hline
Mrk 110 & \makecell{2018-2019 \\2021-2022 }  \\
\hline
Mrk 142 & \makecell{2018-2019 \\2021-2022 }  \\
\hline
Mrk 279 & \makecell{2020-2021 \\2021-2022 }  \\
\hline
Mrk 335 & \makecell{2019-2020 \\2021-2022 }  \\
\hline
Mrk 817 & \makecell{2018-2019 \\2020-2021 \\2021-2022 }  \\
\hline
Mrk 876 & \makecell{2019-2020 }  \\
\hline
Mrk 1044 & \makecell{2018-2019 }  \\
\hline
Mrk 1501 & \makecell{2020-2021 }  \\
\hline
NGC 3516 & \makecell{2019-2020 }  \\
\hline
NGC 7469 & \makecell{2020-2021 }  \\
\hline
PG0052+251 & \makecell{2019-2020 }  \\
\hline
PG0804+761 & \makecell{2019-2020 }  \\
\hline
PG2130+099 & \makecell{2020-2021 \\2021-2022 }  \\
\enddata
\tablecomments{{The final target list for this paper and the seasons that each object was observed in. More detailed lightcurve information is located in the Appendix.}}
\centering
\end{deluxetable}

 \begin{figure*}
 \makebox[\textwidth][c]{\includegraphics[trim=0.5cm 0.5cm 1.5cm 2.5cm, clip = true,width=1.25\textwidth]{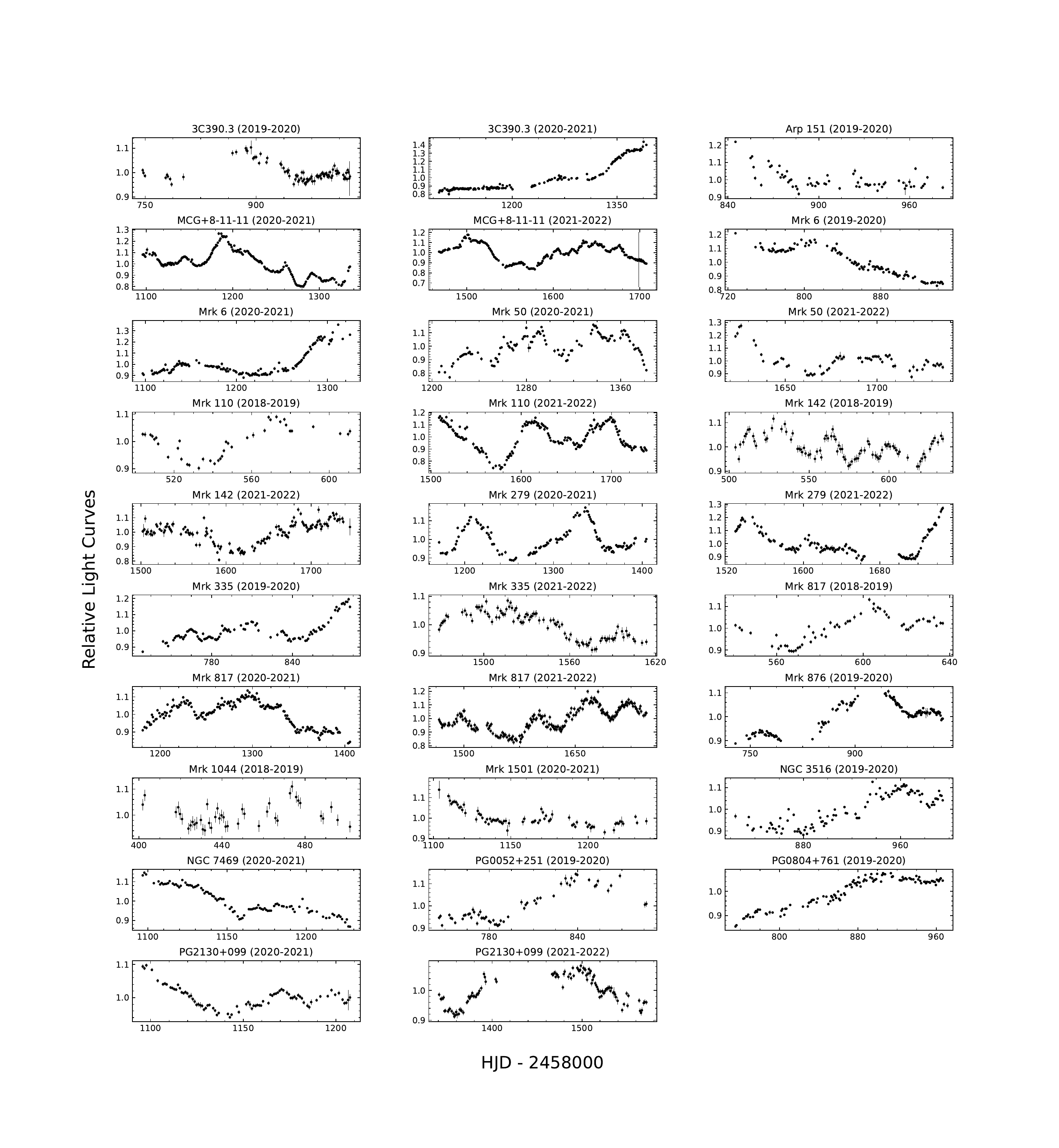}}
 \caption{Observations of the $g$-band for our final sample, listed alphabetically. {The full light curves for every observed band are available in the Appendix.}}
 \label{fig:LC_1}
 \end{figure*}


\section{Analysis and Results}
\label{sec:analysis}
\subsection{Time Lags}
\begin{figure*}
 \makebox[\textwidth][c]{\includegraphics[trim=0.5cm 0.5cm 1.5cm 1.5cm, clip = true,width=1.15\textwidth]{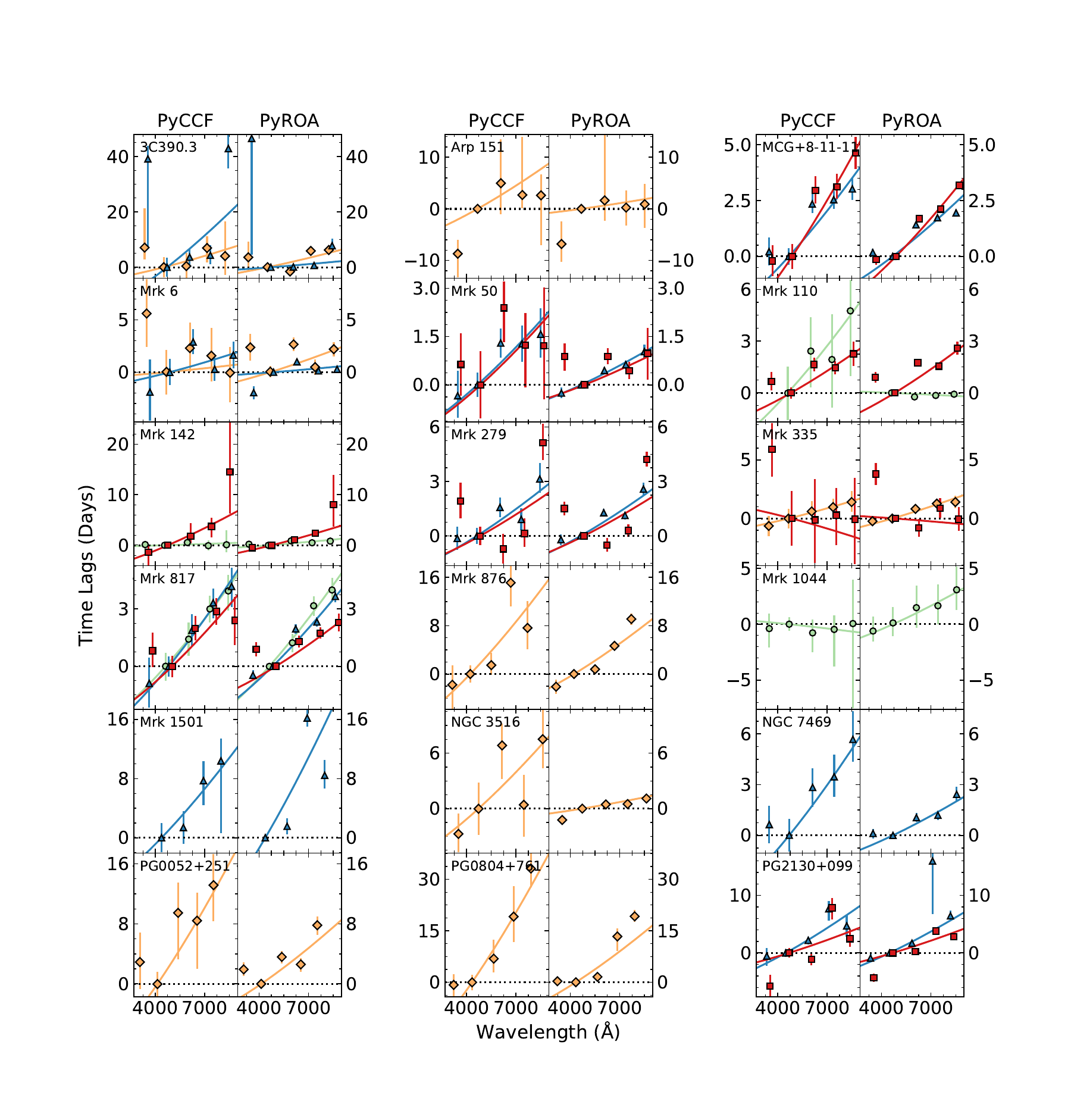}}
 \caption{Measured {PyCCF (left columns) and PyROA (right columns)} time lags. Each color/shape represents a different observation season. {Light green circles are time lags measured from the 2018-2019 observing season, orange diamonds are from the 2019-2020 observations, blue triangles are from the 2020-2021 observing season, and red squares are from the 2021-2022 observing season. An offset of $200\Angstrom$ is applied to each set of time lags after the first on a plot to avoid overlapping.} {Fits to Eq.~\ref{eq:wavlag} with $\beta$ fixed to 4/3 are shown as the solid lines matching the observation year color of the lags. These lags and the value of $\tau_0$ derived from these fits} are given in tables by observation year in the Appendix (Tables \ref{table:year1_lags}, \ref{table:year2_lags}, \ref{table:year3_lags}, and \ref{table:year4_lags}). }
 \label{fig:PyCCF_Lags}
 \end{figure*}

Once the light curves are assembled, we measure time lags between the bands using PyCCF and {PyROA}. For all methods, we use the $g$ band as the reference band due to it having the highest signal-to-noise and best sampling rate for the majority of the objects. PyCCF \citep{2018ascl.soft05032S} calculates the cross correlation function (CCF) \citep{koratkar1991} as implemented by \cite{2004ApJ...613..682P} of two light curves using interpolation to fill in observational gaps. The centroid {of the CCF} {calculated} at 80\% of the {peak correlation coefficient} is used as the time lag between them \citep{koratkar1991a, 2004ApJ...613..682P}. {Flux randomization and random subset selection \citep{ peterson1998} is used to determine the uncertainties on the measured lags.} One limitation of the CCF method is the linear interpolation between data points, which could skew time lags measured from light curves with large gaps between observations. {PyROA \citep{donnan2021} uses a running optimal average to model the driving light curve. PyROA normalizes this model, then shifts and scales it to fit each individual light curve, with Markov Chain Monte Carlo samples to provide uncertainties on parameter estimates. The measured lags for each year of observation are {presented in Tables \ref{table:year1_lags}, \ref{table:year2_lags}, \ref{table:year3_lags}, and \ref{table:year4_lags}}. {Time lag as a function of passband central wavelength for each target is plotted in Fig. \ref{fig:PyCCF_Lags} for both methods.}}

{Assuming} a standard lamppost-like model of an X-ray corona that {irradiates} a geometrically thin, optically thick accretion disk \citep{1973A&A....24..337S}, {geometrical approximations may be used} to predict what the time lags should be for each AGN. Thin disk theory {results in a} temperature profile of 
$T(R) \propto (M\dot{M})^{1/4}R^{-3/4}$, where $T$ is the temperature of the disk, $M$ is the black hole mass, $\dot{M}$ is the {physical} accretion rate of the black hole, and $R$ is the radius of the disk. If we assume the disk is a blackbody, Wien's law then gives $\lambda$~$\propto$~$T^{-1}$. From the lamppost model we assume that time lag $\tau$ should be linked to the radius $R$ of the disk such that $\tau~\sim~R/c$, where $c$ is the speed of light. Combining all of these, we get a relationship between the measured lag and the wavelength of the emitted light:
\begin{equation}
\tau \propto (M\dot{M})^{1/3}T^{-4/3} \propto (M\dot{M})^{1/3}\lambda^{4/3}.
\label{eq:lageq}
\end{equation}

\begin{figure}
 \centering
 \includegraphics[width=0.45\textwidth]{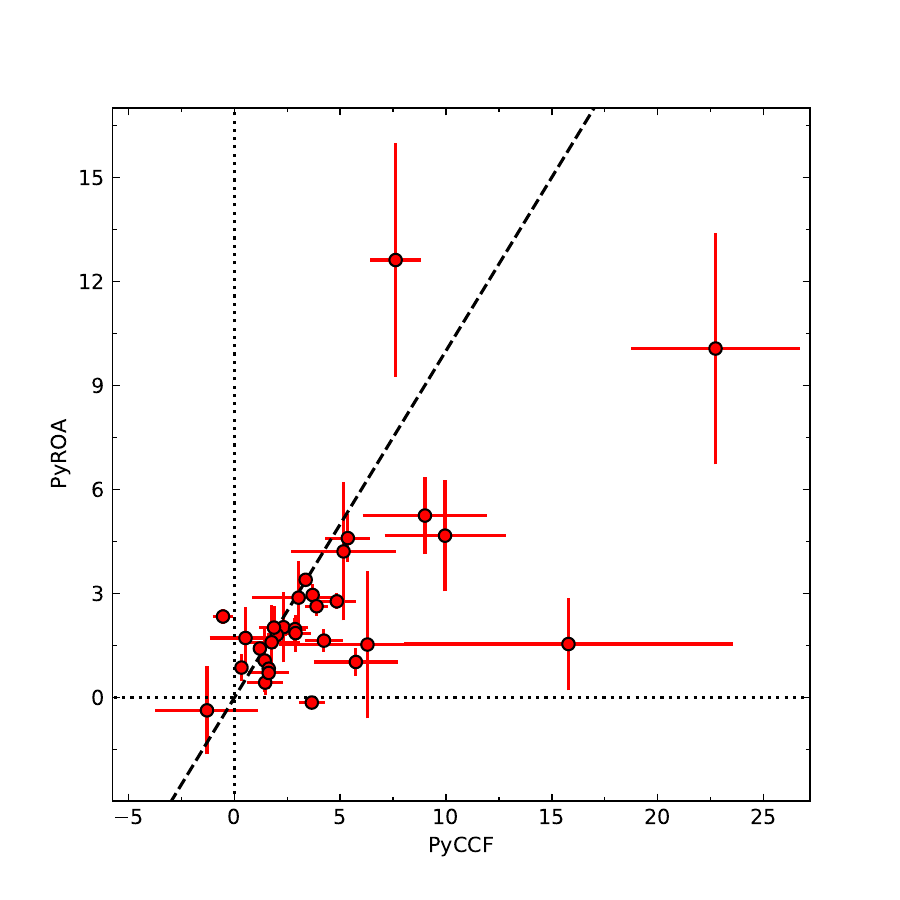}
 \caption{{PyCCF vs. PyROA measured disk sizes in days. Each data point represents one year of observation {for} a single AGN. The dashed black line indicates one-to-one agreement. The vertical and horizontal dotted lines represent when PyCCF or PyROA disk sizes are equal to {zero}, respectively.}}
 \label{fig:Tau0_Comp}
 \end{figure}

{The relationship above predicts that $\tau \propto \lambda^{4/3}$ if the disk behaves in accordance with thin disk theory. We can parameterize this relationship into the equation below,}
\begin{equation}
\tau = \tau_0[(\lambda/\lambda_0)^\beta - 1],
\label{eq:wavlag}
\end{equation}
where $\tau_0$ is the normalization parameter (in units of days), $\lambda$ is the observed wavelength, $\lambda_0$ is the wavelength of the reference band (for this study the $g$ band), and $\beta$ is determined by the relationship of the wavelength to the measured lags. A standard thin disk predicts $\beta$ = 4/3. We test both allowing $\beta$ to be a free parameter and fixing it to 4/3. {These fits are found using the non-linear least squares method employed by the \texttt{curve\_fit} function, available as a part of the publicly available \texttt{scipy} python package \citep{virtanen2020}.} We find that {for many AGN the freely-fitted $\beta$ does not agree with 4/3 or converge to any particular result. This has been observed in several other studies \citep{fausnaugh2018, yu2020, jha2022}. {This may in part be due to influence from the BLR on the $u$ band, which can distort the shape of the underlying powerlaw. We} consider only when $\beta$~=~4/3, and} AGN can then be evaluated as to whether they produce lags in agreement or disagreement with thin disk theory. {We compare the measured $\tau_0$ found with PyCCF and PyROA in Fig. \ref{fig:Tau0_Comp}. {We find that on average the implied disk sizes between each method agree with each other, but in some cases PyCCF finds larger lags.}}

{Some AGN exhibit long-term monotonic increases or decreases throughout the campaign (eg., 3C390.3, Arp~151). It is common in AGN studies to perform a detrending on the data to remove such trends that could bias the time lag measurements \citep{1999PASP..111.1347W, santisteban2020, Miller2023}. However, the exact mechanism to perform the detrending is not uniform, with each AGN usually requiring a unique approach to robustly detrend the data. As a test, we attempt a simple linear fit to each AGN, then subtract out this fit to detrend the data. We measure the time lags afterwards, then compare to the non-detrended data to see if the uncertainties on the time lag measurements have improved. We did not see any substantial improvements, so we elect to continue our analysis on the non-detrended data. It may be that more sophisticated methods of detrending would be required to better account for this long-term variability, which is beyond the scope of this analysis.}


\subsection{Disk Size Comparison With Theory}
{In addition to measuring the implied disk size from time delays of light curves, the expected disk size that emits radiation at wavelength $\lambda_0$ can be calculated from parameters such as SMBH mass {$M$} and the Eddington luminosity {$L_{\mathrm{Edd}}$}.} Following the parameterization of \cite{fausnaugh2016}, we have
\begin{equation}
\tau_0 = \frac{1}{c} \left(X\frac{k \lambda_0}{hc}\right)^{4/3} \left[\left(\frac{GM}{8 \pi \sigma}\right) \left(\frac{L_{\mathrm{Edd}}}{\eta c^2}\right) \left( 3+\kappa \right) \dot{m}_{\rm E} \right]^{1/3},
\label{eq:alpha}
\end{equation}
{where $\eta$ is the radiative efficiency of the conversion from rest mass to radiation, $\kappa$ is the local ratio of internal to external heating sources, {and $\dot{m}_{\rm E}$ is the Eddington ratio, calculated using a bolometric correction of $L_{\rm bol} \sim 9\lambda L_{\lambda}(5100\angstrom)$ \citep{2000ApJ...533..631K}}. $X$ is a factor accounting for how the temperature of the disk relates to the emitted wavelength for a given disk radius.} A major assumption of this parameterization is that the viscous heating within the disk provides roughly the same amount of energy to the disk as the X-rays. {This allows us to define $\eta$~=~0.1 {and $\kappa$~=~1}. A flux-weighted disk gives $X$ = 2.49 \citep{fausnaugh2016}, but when variation of the disk emission is considered this increases to become $X$ = 5.04 \citep{2018MNRAS.473...80T}.} {When using X = 2.49 we find that the measured PyCCF disk sizes are on average 6.5 times larger than predicted, with PyROA disk sizes being 3.8 times greater. However, using 5.04 brings the the average sizes to only being 2.5 times larger for PyCCF and 1.5 times greater for PyROA. {All predicted values of disk size are presented in Table~\ref{table:disksizes}.}

{Eq. \ref{eq:alpha} represents an analytic approximation to accretion disk sizes. Numerical approaches to lag fitting \citep[e.g.][]{mchhardy2018, kammoun21a} take into account additional factors such as inclination, disk color correction factors, disk winds, X-ray source height, and more to fit the measured time lags. These approaches often result in closer agreement between predicted and inferred disk sizes, but require additional parameters that are unknown or are not readily available for the majority of our sample.}}

\subsection{{Flux Calibration}}\label{sec:fluxcal}
To {place our relative photometry onto an absolute scale}, we take the first used epoch of observations for each object of each {observing} season and find all of the detectable stars within that image. The relative photometry is {performed} for that single night of observations, getting the relative fluxes for each star detected in the exposure. The stars are then matched to different stellar magnitude catalogs depending on both the object and filter used. For the majority of our sample, we use APASS data release 10 \citep{henden2018} for the $g$, $r$, and $i$ bands and SDSS data release 18 for the $u$ and $z$ \citep{almeida2023}. We used Pan-STARRS \citep{chamber2016} for some objects that were not covered by either survey. Once the star magnitudes are found, we perform sigma clipping {with a limit of $3\sigma$ to remove outlying stars. We then calculate} the median stellar magnitude for the rest of the conversion {using the zero points for the SDSS filters as determined by \cite{fukugita1996}. We divide this result by the corresponding relative flux for that night of observation to normalize the flux to the mean of the entire light curve.} Finally, we apply a {geometric} correction for redshift and deredden using Cardelli's {extinction} law \citep{cadelli1989_extinction} with E(B-V) values found from \cite{schlafly2011}. Every object in our sample had stars in the field of view with {corresponding} APASS catalogue data for the $g$, $r$, and $i$ bands. However, coverage was not as complete for SDSS $u$- and $z$-band, so some objects were not able to have these relative fluxes {calibrated to absolute fluxes. The objects that had all five bands available are 3C390.3, Arp~151, MCG+8-11-11, Mrk~50, Mrk~110, Mrk~142, Mrk~335, Mrk~817, Mrk~1044, NGC~7469, PG0052+251, and PG2130+099}.

\section{Discussion}
\label{sec:discussion}
\subsection{Continuum Time Lag Analysis}
{The AGN in our sample show strong variability and are observed frequently enough to measure robust time lags. We attribute these results to high-cadence (1-3~days between observations) monitoring with an observational campaign duration of at least 90~days per object. These parameters appear ideal for characterizing the variability of AGN with SMBH masses between $10^{7-8}~\textrm{M}_\odot$. Most robotic ground-based observatories are equipped with broadband optical filters that can observe at these cadences rather trivially, making them ideal for observing AGN in this mass regime and constraining the outer edges of their reprocessing regions.} 

{In general, we find that {inferred disk sizes} {{for PyCCF and PyROA}} are consistent with one another within {the} uncertainties, although PyCCF disk sizes tend to be slightly larger {when they do disagree (see Fig.~\ref{fig:Tau0_Comp}). Measured} time lags are compared to the predicted time lags from thin disk theory. We find, like many studies have previously found, that inferred disk sizes are larger than predicted in the simplest analytical form of the lamp-post model.} 


\subsection{Luminosity Dependence on Lags}
\label{Sec:Expanded_Sample}
The measured time lags can also be compared to physical properties of the AGN, such as its mass and luminosity. 
{As discussed in the Introduction, both the accretion disk model and the BLR-dominated model lead to the prediction that $\tau \propto L^{1/2}$. Previous work has made the assumption of a constant bolometric correction, leading to a slightly different dependence, and we present a similar approach in the Appendix.}



{We can test this prediction by plotting disk size versus luminosity. We approximate the 5100$\Angstrom$~luminosity for our sample by linearly interpolating between the redshift/dust-corrected $g$- and $r$-band fluxes. {We take the standard deviation of the light curve as the average uncertainty for both the $g-$ and $-r$ light curves and combine them in quadrature to get the uncertainty at 5100$\Angstrom$. We subtract host galaxy fluxes that were measured from high-resolution HST images through the Zowada photometric aperture size following the methods of \cite{bentz2009, bentz2013}. The redshifts, masses, observed luminosities, host galaxy luminosities, and host-subtracted luminosities are presented in Table~\ref{table:AGN_facts}.} We also include objects for which {host-galaxy correction} is unavailable {as a separate {dataset; these} fluxes should be treated as upper limits on the AGN's brightness.} }

{In order to explore as wide a range of $M, \dot{M},$ and $L_{5100}$ as possible, we include {measurements from} several other studies {in} our analysis.} We only {incorporate} studies that {have} measured time lags, masses, and 5100$\Angstrom$ luminosity measurements for all sources. This includes Fairall~9 \citep{santisteban2020}, MCG+8-11-11 \citep{fausnaugh2018}, Mrk~110 \citep{vincentelli2021}, Mrk~142 \citep{cackett2020}, Mrk~335 \citep{kara2023}, Mrk~509 \citep{edelson2019}, Mrk~817 \citep{kara2021}, Mrk~876 \citep{Miller2023}, NGC~2617 \citep{fausnaugh2018}, NGC~4151 \citep{edelson2019}, NGC~4593 \citep{cackett2018}, NGC~5548 \citep{fausnaugh2016}, and NGC~7469 \citep{vincentelli2023}. {This sample represents 13~local, luminous, and well studied AGN with wavelength coverage of the entire optical continuum and often includes simultaneous UV/X-ray observations.} {We {take the reported luminosities} from this sample from their papers. If the $5100\Angstrom$ luminosity is not reported, we take measurements from the AGN black hole mass database \citep{bentz2015}} which have been host-galaxy subtracted but may not have been taken synchronously with their measured time lags. We refer to this sample as the intensive broadband reverberation mapping (IBRM) sample henceforth. {We also include a sample of 38~AGN from the ``Core" sample of time lags determined from the Zwicky Transient Facility (ZTF) \citep{guo2022} that is referred to as the Guo dataset. {We adopt 0.05~dex uncertainties on Guo dataset's luminosities to match the values they use in their study, and use the same uncertainty for all objects where the uncertainty of the 5100$\Angstrom$ luminosity is not reported. 

{{As to not double-count a specific AGN during the same observing period, we do not include Zowada disk measurements where Zowada observations contributed to the final light curve of that study (see Sec.~\ref{sec:methods} for a complete list of these studies).} As such, we do not use Mrk~110 (2018-2019), Mrk~142 (2018-2019), {Mrk~335 (2019-2020)} and Mrk~817 (2018-2019).} {Measurements that will be used in future studies but that are not currently published are included}. {We do not average together multiple years of observation for this analysis. Each year of observation represents a discrete accretion rate and size measurement of the reprocessing region. We utilize each year of time lags for each object as its own data point.} This brings the Zowada dataset to 25 sets of time lags {and the total dataset to {76~sets}} of time lags initially. Since the majority of these sources use PyCCF lags for their primary analyses, we likewise use our PyCCF lags and derived disk sizes for all further analysis.

Each study's lags are adjusted for redshift if necessary, then fit {with \texttt{curve\_fit}} using their respective reference bands as $\lambda_0$ and fixing $\beta$ to be 4/3. {We then extrapolate to the disk size at 5100$\Angstrom$ following that 4/3 relation.} If a study excludes {any specific bands} from their analysis, we do so as well. Finally, we exclude any objects that result in a $\tau_0$ consistent with 0 {within $1\sigma$ uncertainty}. Both the IBRM and Guo samples all pass this final filter, with only {three~Zowada targets failing {(Mrk~6 (2019-2020), Mrk~335 (2021-2022) and Mrk 1044 (2018-2019))}, resulting in a final sample of {73~AGN {disk size--luminosity} measurements.}} 

\subsection{Luminosity and Disk Size}

\begin{figure*}
 \centering
 \includegraphics[trim=0.5cm 0cm 1cm 1cm, clip = true,width=0.99\textwidth]{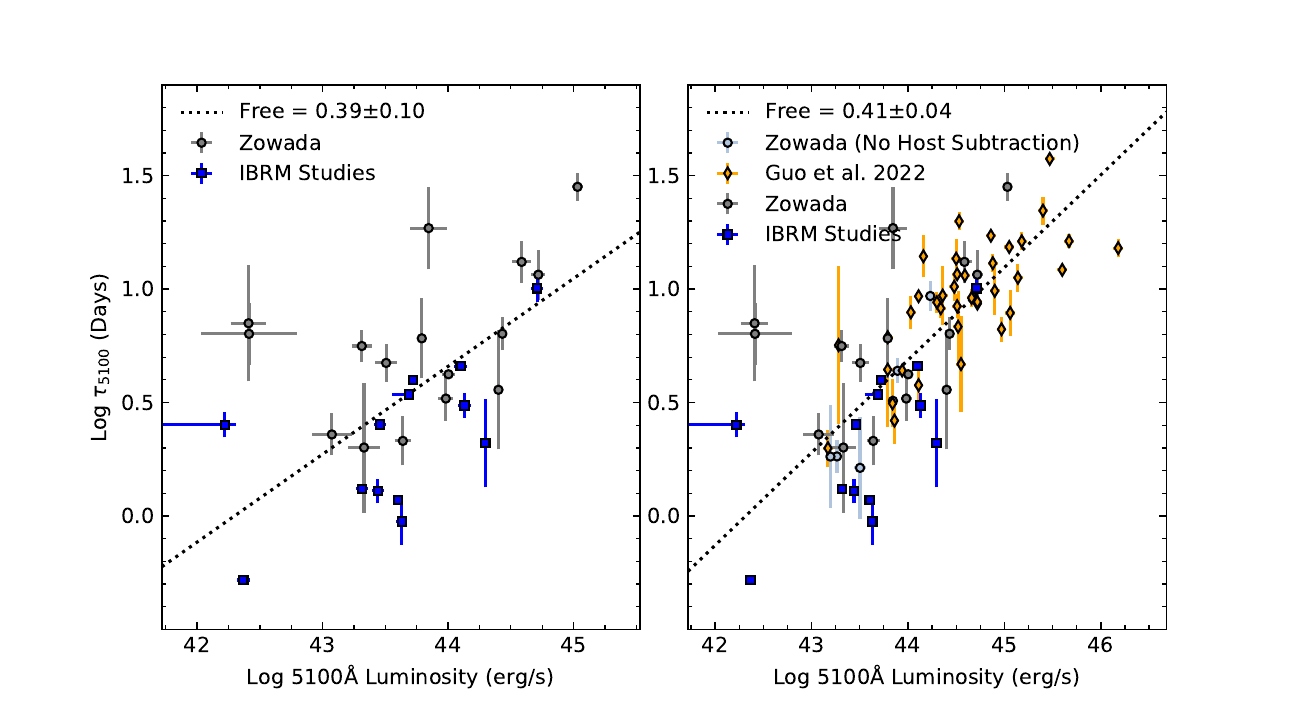}
 \caption{The relationship between log AGN luminosity and measured disk size. {Gray circles and light-gray} circles are Zowada objects with and without host subtraction, respectively. Blue squares are the {results from} IBRM studies, and orange diamonds are from \cite{guo2022}'s ZTF study. {The dotted line is a linear fit to the data using linmix \citep{kelly2007}. We find that both datasets prefer the BLR interpretation of $L_{5100}^{0.5}$, however there exists {sufficient scatter in the relation that a thin-disk interpretation cannot be ruled out.}}}
 \label{fig:lum_plot_both}
 \end{figure*}

{To probe the effect {that uncorrected host-galaxy contributions may have} on the sample, we perform our analysis in two parts. We first consider only AGN that have host-galaxy-modeled subtraction available and {extended wavelength coverage of the optical bands}, then expand the analysis to include the entire dataset. The former sample includes only the Zowada and IBRM studies, resulting in a slope that is less constrained with $L_{5100}^{0.39\pm0.10}$. When fitting the entire dataset, we find $L_{5100}^{0.41\pm0.04}$. Both results are in agreement with previous studies \citep{netzer2022, guo2022, montano2022}. While close to agreement, the fits are slightly shallower for both data sets than the expected value of $L_{5100}^{0.5}$. {More} observations of low ($<10^{42}$~$\mathrm{erg\ s^{-1}}$) and high ($>10^{45}$~$\mathrm{erg\ s^{-1}}$) luminosity AGN are needed to more reliably constrain this relation. With only four~$\tau_0$ measurements {below $10^{43}$}~$\mathrm{erg\ s^{-1}}$ and only a single measurement {for $\tau_0$} above $10^{46}$~$\mathrm{erg\ s^{-1}}$, these areas remain our most crucial parameter space for further study. When we exclude these measurements from the fit, the slope changes to $0.54\pm0.12$ and $0.47\pm0.04$ for the host-subtracted and full datasets, respectively, in full agreement with predictions.}

{Proper host galaxy subtraction is vital to avoid overestimating the AGN's true luminosity. This is especially important at low luminosities, where the host galaxy is typically a larger fraction of the total flux. Without this subtraction the relation could be artificially flattened, leading to incorrect results.} {Previous studies of the BLR~radius compared to host luminosity \citep[e.g.][]{kaspi2005} were skewed by failing to account for the contribution from host-galaxy flux \citep{bentz2006, bentz2009}.} However, due to the intense modeling and observations required for such removal, this has not been done for a large amount of objects, although it has been improving in recent years. Several objects in the Guo dataset have their host galaxy contributions removed through spectral fitting, but there are some that do not have this removal. This effect was found to be minimal in previous studies comparing these catalogs and finds that the effect is less pronounced in higher luminosity AGN \citep{rakshit2020}. In addition, this is not a simultaneous luminosity measurement with the fluxes, nor is it scaled to match the aperture of ZTF.

It is also possible that the nature of optical AGN variability changes with increasing mass and luminosity. {There exist many formulations for the self-gravitational radius of an accretion disk. One prediction suggests an upper limit of $\sim$12~light~days \citep{lobban2022}.} {If this is correct and interband continuum delays are disk-dominated, then we would expect lags to increase with luminosity until the disk hits {its} size limit, where then the implied size should flatten out with increasing luminosity,} {Our results could indicate that for lower masses and luminosities, the variability pattern is more indicative of thin disk emission with some BLR contribution, {{similar to} what we observe with the host-subtracted dataset in Fig.~\ref{fig:lum_plot_both}}. However, different formulations of the disk result in different predictions for the maximum disk radius and may not account for different accretion regimes for both lower and higher accretion rates \citep[e.g.][]{laor1989, gammie2001}. As such, no definitive conclusions can be drawn without firmly establishing disk sizes across the entire scale of AGN accretion rates.}

{ The luminosity-dependence of continuum lags can be probed further with}
{the advent of large all-sky surveys such as the Legacy Survey of Space and Time, as an extremely large volume of AGN can be examined. Expanding the number of AGN examined with luminosity below $<10^{42}$~$\mathrm{erg\ s^{-1}}$ and above $>10^{45}$~$\mathrm{erg\ s^{-1}}$ is necessary to completely verify this claim. There may also be differences in super-Eddington AGN, which has been found to exist in BLR RM studies \citep{du2018, yao2024}.}

\subsection{Notes On Specific AGN}

\subsubsection*{Mrk 335}
{Mrk~335 has been monitored recently by both \cite{komossa2020} and \cite{kara2023}. Our 2019-2020 observations of Mrk~335 are a part of the latter's analysis, and both studies report that Mrk~335 is emerging from a low state around the end of that season's observations. This rise could be explained by a weakening partial obscurer or by a rapid inflow of material that triggers both X-ray bursts and a large optical increase. We also present data two years later (the 2021-2022 observing season), finding the variability to be weaker (see Tables~\ref{table:year2_lt}~and~\ref{table:year4_lt}) and with a more bumpy/nonuniform CCF. This {absence} of variability resulted in a derived continuum emitting size consistent with zero, and as such was not included in Fig.~\ref{fig:lum_plot_both}.}

\subsubsection*{Mrk 817}
Mrk~817 was recently observed {by} the AGN~STORM~2 \citep{kara2021} campaign. Their time lags are measured relative to the $UVW2$ band at 1928$\Angstrom$, but when {referenced} with respect to the $g$~band {their time lags agree with each set of Zowada lags across all 3~years.} Zowada observations from 2020 onwards are included in the AGN STORM 2 campaign. { When scaled to $5100\Angstrom$, we find that the disk size is consistent with the 2021-2022 observing season, and relatively close to the sizes of the other observing seasons but disagrees within $1\sigma$ uncertainties. This may be because of our poorer sampling compared to their more complete observation cadence and broader wavelength coverage, or perhaps due to} longer term variations within their extended dataset. The full optical RM analysis portion of the AGN~STORM~2 campaign is not yet complete, so {further discussion {of this point} will be reserved for a future paper.}

\subsubsection*{Mrk 876}
Mrk~876 was observed recently in concert with the Las Cumbres Observatory and Liverpool Telescope in combination with Zowada observations from the 2018-2019 observation cycle \citep{Miller2023}. Unfortunately, this light curve was removed from the final data set due to a lack of variability in the Zowada data alone. However, it was observed again in the 2019-2020 dataset and shows remarkable variability. The PyCCF time lags and measured disk sizes are in agreement {between this previous study and ours, including an excess in the $i$-band lags.} {A major difference between the analysis {presented} here and the previous study is that {in the latter, the variability data were first detrended}. A moving boxcar average was employed to remove long term trends, and it was found that a boxcar width of 100~days was {required}. It is interesting that, despite not detrending, a similar result is found. This could indicate that the longer term variations found within the three~year dataset were not manifested within the single year of observation we have for this current analysis.} 


\subsubsection*{PG2130+099}
This object was observed twice by Zowada in the 2020-2021 and 2021-2022 observation seasons, finding results that generally agree {within the uncertainties} for the lags and disk sizes. This object has had a unique past with both BLR RM \citep{grier2008, grier2013, hu2020, yao2024} and {continuum RM}. \cite{jha2022} observed this object using SDSS $gri$ filters and found that its disk measurement agrees with thin disk predictions. Meanwhile, \cite{fian2022} uses special filters to circumvent potential H$\alpha$ contamination and finds using multiple different lag measurement methods that the disk size exceeds thin disk predictions by a factor of 2-6 depending on the method. {We contributed our data towards another study \citep{miller25} observing PG2130+099 again with higher cadence observations. That study found evidence for {longer lags} with increasing luminosity.}

Our analysis shows agreement between the measured disk sizes at the $g$-band effective wavelength as calculated by \cite{jha2022} for both seasons of our observations, although the 2021-2022 observation season {was impacted by an observation} gap and is less robust. {We detect an $i$-band excess lag similar to what is observed in Mrk~876. This is likely due to H$\alpha$ emission, which is expected at PG2130+099's redshift of 0.0629.}

\section{Conclusions}
\label{sec:conclusions}
In this work we analyze 29~sets of $ugriz$ light curves from a total of 18~AGN across four~years of observations. These light curves are found to be variable enough to recover robust time lags. We measure these lags with both PyCCF and PyROA, and for most objects these lags {and measured disk sizes} agree within {the uncertainties}. 

As has been found with previous studies, we find the measured disk sizes to be consistently larger than predicted using standard assumptions of disk geometry when considering only a flux-weighted mean radius geometry. If instead {disk variability is included in the calculation (resulting in a value for X~=~$5.04$)}, we find much closer results, {especially for disk sizes determined with PyROA.} This could indicate that the larger than expected disk size problem found in many previous studies is {due to not accounting for this variability, although notably it does not completely fix the discrepancy for either PyROA or PyCCF.} {More {sophisticated disk-dominated} models \citep[ex.][]{sun2020b, kammoun2023} can fit time lags without a discrepancy.}

 
{For the 18~AGN analyzed, we interpolate the $g-$ and $r-$band observations to recover ${L}_{5100}$.} Using archival HST data and galaxy modeling, we remove host galaxy {starlight contribution} from ${14}$ AGN. 
{Using the AGN lags presented in this paper and IBRM studies, as well as when extending this analysis to objects without host galaxy correction and with less wavelength {coverage}, we find that the best-fit {relationships of $\tau_{5100} \propto L_{5100}^{0.39\pm0.10}$ for the host-subtracted dataset and $\tau_{5100} \propto L_{5100}^{0.41\pm0.04}$ for our entire dataset agree with BLR variability.} This result is consistent with previous studies \citep{netzer2022, guo2022, montano2022},}
{ and consistent with both the disk-dominated and BLR-dominated models for continuum variability.}

This data set extends the analysis of previous papers and is consistent with previous results, {with specific attention drawn to how host-galaxy subtraction can affect the observed relationships.} {The host-subtracted and full continuum dataset does not span the same mass and luminosity range compared to the entire dataset, which could impact the slope measurement between these two parameters. It also raises some concerns about how this effect may evolve with increasing AGN mass and luminosity}. If the self-gravity limit of 12~light~days \citep{lobban2022} is correct, then this could indicate a transition {from disk-dominated variability to one of BLR dominance. Across all samples, there is a dearth of objects at the high and low end of the M-L plot {(Fig.~\ref{fig:massvlum})}. For example, there are only a handful of objects with a measured luminosity greater than $10^{45}$~$\mathrm{erg\ s^{-1}}$, and none with complete optical continuum coverage. There is also a large parameter space to be explored in the form of intermediate mass black hole AGN \citep{montano2022, zuo2024}. More samples spanning an even wider range of AGN masses and luminosities, {as well as careful correction for the host galaxy, are} necessary to resolve this discrepancy.} 

Time-domain survey studies of AGN can provide an extremely large amount of light curves, but currently suffer from a {coarse observational cadences and limited wavelength coverage.} However, the next generation of time-domain surveys such as the Vera C. Rubin Observatory's Legacy Survey of Space and Time will monitor the southern sky with $ugriz$ observations at a roughly 3~day cadence {in certain observational programs. Combining these data with observations from observatories like Zowada will produce light curves capable of recovering robust lag measurements}. 
\section{Acknowledgments}
{We graciously thank the anonymous referee for their detailed and thoughtful comments that have substantially improved this manuscript.} MCB gratefully acknowledges support from the NSF through {grants AST-2009230 and AST-2407802. JAM and EMC gratefully acknowledge support from the National Science Foundation through AST-1909199.}

\appendix
\label{sec:appendix}
{Here we present all tables and the complete light curves for each object in our sample. AGN are listed alphabetically in observation order. The light curves are presented in relative count format. CCFs with their measured lags are presented next to each light curve, with the $g$ band always being the reference band. The solid black vertical line is the measured lag, the dotted line {represents} the uncertainties, and the dashed line is {set to zero} for reference. }

{We also include the determination of an empirical relationship between SMBH mass and luminosity, similar to what previous studies have found. In addition to the IBRM and Guo datasets, we also include \cite{2017ApJ...836..186J}'s AGN masses and luminosities to extend our parameter space to the largest range probed thus far. {To avoid repeat counting AGN in the fit, studies of the same source have their luminosities' averaged together.} The fitted relationship is shown in Fig.~\ref{fig:massvlum} {as the solid black line. Note that this plot shows all values of our sample, not the averages used in the fit. Thus, any points at the same mass are averaged together before fitting.}} 

{{For our sample, we use masses taken from the Black Hole Mass Database \citep{bentz2015}, with the only exception being Mrk~1044 which we take from \cite{wang2014}.} For IBRM masses, we use the masses and uncertainties as presented in \cite{bentz2015} assuming a scale factor $<f>=4.3$ {\citep{grier2013a}}. If the uncertainty is not known, we adopt the uncertainty of 0.31~dex as has been shown through scaling relations \citep{shen2024}. {Following previous studies, we fit a power law between $M$ and {$L_{5100}$}, finding a {slope of $1.24 \pm 0.10$.} We then find the inverse of this relationship, such that $M \propto L_{5100}^{0.81\pm0.07}$. This relationship is substituted on the right side of Eq.~\ref{eq:lageq} to remove the mass term. Mass accretion is directly proportional to luminosity, and by replacing that term with luminosity, Eq.~\ref{eq:lageq} would simplify to $\tau_0 \propto L_{5100}^{0.60 \pm0.02}$ as the prediction for accretion-disk variability.}}
\clearpage
\begin{rotatetable}
\movetableright=0.25cm
\begin{deluxetable*}{lccccccccc}
\tablecolumns{7}
\tablecaption{2018-2019 Observations
\label{table:year1_lt}}
\tablehead{
\colhead{} & \multicolumn{4}{c}{Lightcurve Information}  & \multicolumn{5}{c}{F$_{var} (\%)$} \\
Object & Start Date & End Date & Length (Days) & Average Cadence (Days) & $u$ & $g$ & $r$ & $i$ & $z$ }
\startdata
Mrk 110 & 503.84 & 610.71  & 106.87  & 2.41 & -- & 0.0533 $\pm 0.0012$ & 0.0420 $\pm 0.0016$ & 0.0396 $\pm 0.0010$ & 0.0387 $\pm 0.0019$ \\
Mrk 142 & 503.90 & 633.67  & 129.77  & 1.76 & 0.0783 $\pm 0.0056$ & 0.0402 $\pm 0.0017$ & 0.0281 $\pm 0.0013$ & 0.0266 $\pm 0.0020$ & 0.0166 $\pm 0.0031$ \\
Mrk 817 & 541.00 & 636.78  & 95.78  & 1.84 & -- & 0.0591 $\pm 0.0011$ & 0.0454 $\pm 0.0013$ & 0.0416 $\pm 0.0007$ & 0.0318 $\pm 0.0008$ \\
Mrk 1044 & 401.83 & 501.66  & 99.83  & 2.22 & -- & 0.0348 $\pm 0.0037$ & 0.0384 $\pm 0.0031$ & 0.0301 $\pm 0.0026$ & 0.0319 $\pm 0.0026$ \\
\enddata
\tablecomments{Lightcurve Information is for the $g$-band lightcurves of each object. Dates are listed as HJD - 2458000. F$_{var}$ is the excess variability for each band.}
\centering
\end{deluxetable*}
\end{rotatetable}

\begin{rotatetable}
\movetableright=0.25cm
\begin{deluxetable*}{lccccccccc}
\tablecolumns{7}
\tablecaption{2019-2020 Observations
\label{table:year2_lt}}
\tablehead{
\colhead{} & \multicolumn{4}{c}{Lightcurve Information}  & \multicolumn{5}{c}{F$_{var} (\%)$} \\
Object & Start Date & End Date & Length (Days) & Average Cadence (Days) & $u$ & $g$ & $r$ & $i$ & $z$ }
\startdata
3C390.3 & 746.71 & 1026.79  & 280.08  & 2.61 & 0.0865 $\pm 0.0064$ & 0.0326 $\pm 0.0015$ & 0.0223 $\pm 0.0008$ & 0.0225 $\pm 0.0009$ & 0.0211 $\pm 0.0022$ \\
Arp 151 & 844.98 & 981.85  & 136.87  & 2.09 & 0.1470 $\pm 0.0267$ & 0.0525 $\pm 0.0012$ & 0.0436 $\pm 0.0008$ & 0.0254 $\pm 0.0010$ & 0.0252 $\pm 0.0022$ \\
Mrk 6 & 727.97 & 944.71  & 216.74  & 1.99 & 0.1487 $\pm 0.0026$ & 0.0992 $\pm 0.0009$ & 0.0927 $\pm 0.0007$ & 0.0810 $\pm 0.0008$ & 0.0823 $\pm 0.0008$ \\
Mrk 335 & 728.95 & 882.62  & 153.67  & 1.78 & 0.1001 $\pm 0.0022$ & 0.0653 $\pm 0.0009$ & 0.0466 $\pm 0.0008$ & 0.0424 $\pm 0.0007$ & 0.0330 $\pm 0.0010$ \\
Mrk 876 & 728.67 & 1025.68  & 297.01  & 2.11 & 0.0478 $\pm 0.0028$ & 0.0540 $\pm 0.0005$ & 0.0456 $\pm 0.0005$ & 0.0329 $\pm 0.0006$ & 0.0273 $\pm 0.0018$ \\
NGC 3516 & 824.00 & 994.80  & 170.80  & 1.85 & 0.2417 $\pm 0.0014$ & 0.0735 $\pm 0.0006$ & 0.0499 $\pm 0.0008$ & 0.0347 $\pm 0.0006$ & 0.0309 $\pm 0.0010$ \\
PG0052+251 & 745.94 & 886.62  & 140.68  & 2.23 & 0.0789 $\pm 0.0030$ & 0.0723 $\pm 0.0013$ & 0.0642 $\pm 0.0011$ & 0.0323 $\pm 0.0011$ & 0.0423 $\pm 0.0022$ \\
PG0804+761 & 754.96 & 966.67  & 211.71  & 1.90 & 0.0696 $\pm 0.0012$ & 0.0593 $\pm 0.0004$ & 0.0590 $\pm 0.0004$ & 0.0440 $\pm 0.0004$ & 0.0525 $\pm 0.0008$ \\
\enddata
\tablecomments{Lightcurve Information is for the $g$-band lightcurves of each object. Dates are listed as HJD - 2458000. F$_{var}$ is the excess variability for each band.}
\centering
\end{deluxetable*}
\end{rotatetable}

\begin{rotatetable}
\movetableright=0.25cm
\begin{deluxetable*}{lccccccccc}
\tablecolumns{7}
\tablecaption{2020-2021 Observations
\label{table:year3_lt}}
\tablehead{
\colhead{} & \multicolumn{4}{c}{Lightcurve Information}  & \multicolumn{5}{c}{F$_{var} (\%)$} \\
Object & Start Date & End Date & Length (Days) & Average Cadence (Days) & $u$ & $g$ & $r$ & $i$ & $z$ }
\startdata
3C390.3 & 1095.72 & 1391.80  & 296.08  & 1.95 & 0.2956 $\pm 0.0054$ & 0.1782 $\pm 0.0009$ & 0.1193 $\pm 0.0007$ & 0.1090 $\pm 0.0007$ & 0.1043 $\pm 0.0014$ \\
MCG+8-11-11 & 1095.91 & 1335.64  & 239.73  & 1.58 & 0.1465 $\pm 0.0033$ & 0.1109 $\pm 0.0005$ & 0.0727 $\pm 0.0004$ & 0.0797 $\pm 0.0004$ & 0.0662 $\pm 0.0005$ \\
Mrk 6 & 1096.95 & 1324.75  & 227.80  & 1.80 & 0.1657 $\pm 0.0075$ & 0.1040 $\pm 0.0010$ & 0.0899 $\pm 0.0006$ & 0.0873 $\pm 0.0006$ & 0.0694 $\pm 0.0009$ \\
Mrk 50 & 1206.00 & 1381.72  & 175.72  & 1.78 & 0.1846 $\pm 0.0054$ & 0.0870 $\pm 0.0010$ & 0.0544 $\pm 0.0009$ & 0.0483 $\pm 0.0013$ & 0.0358 $\pm 0.0021$ \\
Mrk 279 & 1170.98 & 1404.74  & 233.76  & 1.87 & 0.1627 $\pm 0.0043$ & 0.0679 $\pm 0.0007$ & 0.0441 $\pm 0.0008$ & 0.0383 $\pm 0.0009$ & 0.0318 $\pm 0.0008$ \\
Mrk 817 & 1180.98 & 1405.68  & 224.70  & 1.60 & 0.0834 $\pm 0.0034$ & 0.0707 $\pm 0.0006$ & 0.0569 $\pm 0.0005$ & 0.0543 $\pm 0.0005$ & 0.0395 $\pm 0.0007$ \\
Mrk 1501 & 1103.71 & 1237.61  & 133.90  & 1.97 & -- & 0.0392 $\pm 0.0017$ & 0.0354 $\pm 0.0013$ & 0.0434 $\pm 0.0012$ & 0.0402 $\pm 0.0045$ \\
NGC 7469 & 1096.76 & 1227.61  & 130.85  & 1.67 & 0.1298 $\pm 0.0013$ & 0.0692 $\pm 0.0003$ & 0.0451 $\pm 0.0003$ & 0.0426 $\pm 0.0003$ & 0.0345 $\pm 0.0005$ \\
PG2130+099 & 1095.78 & 1207.60  & 111.82  & 1.69 & 0.0293 $\pm 0.0051$ & 0.0309 $\pm 0.0007$ & 0.0302 $\pm 0.0007$ & 0.0259 $\pm 0.0006$ & 0.0247 $\pm 0.0018$ \\
\enddata
\tablecomments{Lightcurve Information is for the $g$-band lightcurves of each object. Dates are listed as HJD - 2458000. F$_{var}$ is the excess variability for each band.}
\centering
\end{deluxetable*}
\end{rotatetable}

\begin{rotatetable}
\movetableright=0.25cm
\begin{deluxetable*}{lccccccccc}
\tablecolumns{7}
\tablecaption{2021-2022 Observations
\label{table:year4_lt}}
\tablehead{
\colhead{} & \multicolumn{4}{c}{Lightcurve Information}  & \multicolumn{5}{c}{F$_{var} (\%)$} \\
Object & Start Date & End Date & Length (Days) & Average Cadence (Days) & $u$ & $g$ & $r$ & $i$ & $z$ }
\startdata
MCG+8-11-11 & 1467.90 & 1707.64  & 239.74  & 1.60 & 0.1177 $\pm 0.0024$ & 0.0810 $\pm 0.0015$ & 0.0527 $\pm 0.0006$ & 0.0562 $\pm 0.0005$ & 0.0463 $\pm 0.0007$ \\
Mrk 50 & 1622.90 & 1735.76  & 112.86  & 1.87 & 0.2531 $\pm 0.0138$ & 0.0795 $\pm 0.0018$ & 0.0482 $\pm 0.0017$ & 0.0430 $\pm 0.0019$ & 0.0424 $\pm 0.0032$ \\
Mrk 110 & 1508.92 & 1738.65  & 229.73  & 1.65 & 0.1572 $\pm 0.0034$ & 0.1052 $\pm 0.0009$ & 0.0606 $\pm 0.0007$ & 0.0711 $\pm 0.0007$ & 0.0568 $\pm 0.0014$ \\
Mrk 142 & 1501.98 & 1745.66  & 243.68  & 1.84 & 0.1835 $\pm 0.0084$ & 0.0712 $\pm 0.0013$ & 0.0510 $\pm 0.0009$ & 0.0414 $\pm 0.0012$ & 0.0635 $\pm 0.0028$ \\
Mrk 279 & 1529.00 & 1745.79  & 216.79  & 1.86 & 0.1595 $\pm 0.0070$ & 0.0931 $\pm 0.0011$ & 0.0484 $\pm 0.0011$ & 0.0444 $\pm 0.0010$ & 0.0371 $\pm 0.0014$ \\
Mrk 335 & 1468.78 & 1613.59  & 144.81  & 1.75 & 0.0630 $\pm 0.0036$ & 0.0448 $\pm 0.0011$ & 0.0334 $\pm 0.0011$ & 0.0399 $\pm 0.0015$ & 0.0405 $\pm 0.0014$ \\
Mrk 817 & 1466.63 & 1745.71  & 279.08  & 1.65 & 0.1183 $\pm 0.0032$ & 0.0823 $\pm 0.0011$ & 0.0709 $\pm 0.0006$ & 0.0630 $\pm 0.0006$ & 0.0487 $\pm 0.0013$ \\
PG2130+099 & 1340.94 & 1571.59  & 230.65  & 2.04 & 0.0604 $\pm 0.0036$ & 0.0469 $\pm 0.0008$ & 0.0416 $\pm 0.0007$ & 0.0327 $\pm 0.0008$ & 0.0324 $\pm 0.0015$ \\
\enddata
\tablecomments{Lightcurve Information is for the $g$-band lightcurves of each object. Dates are listed as HJD - 2458000. F$_{var}$ is the excess variability for each band.}
\centering
\end{deluxetable*}
\end{rotatetable}

\begin{deluxetable*}{lccccccc}
\tablecolumns{8}
\tablecaption{2018-2019 Time Lags
\label{table:year1_lags}}
\tablehead{
Object & Method & $u$ & $g$ & $r$ & $i$ & $z$ & $\tau_0$ }
\startdata
Mrk 110 & PyCCF 
 & $--$
 & $-0.02^{+1.55}_{-1.57}$
 & $2.42^{+1.96}_{-2.09}$
 & $1.93^{+2.64}_{-2.78}$
 & $4.76^{+1.80}_{-3.78}$
 & $3.66 \pm 0.61$
\\
 & PyROA 
 & $--$
 & $-0.00 \pm 0.05$
 & $-0.23^{+0.15}_{-0.12}$
 & $-0.15^{+0.15}_{-0.12}$
 & $-0.08^{+0.16}_{-0.13}$
 & $-0.14 \pm 0.08$
\\
\addlinespace
Mrk 142 & PyCCF 
 & $0.14^{+0.69}_{-0.83}$
 & $-0.00 \pm 0.37$
 & $0.55^{+0.57}_{-0.58}$
 & $-0.08^{+1.08}_{-0.96}$
 & $0.10^{+2.81}_{-1.50}$
 & $0.33 \pm 0.33$
\\
 & PyROA 
 & $0.21^{+0.28}_{-0.30}$
 & $-0.01^{+0.17}_{-0.16}$
 & $0.88^{+0.22}_{-0.23}$
 & $0.50^{+0.34}_{-0.38}$
 & $0.82^{+0.71}_{-0.66}$
 & $0.87 \pm 0.40$
\\
\addlinespace
Mrk 817 & PyCCF 
 & $--$
 & $0.00^{+0.71}_{-0.76}$
 & $1.42 \pm 0.87$
 & $3.00^{+0.71}_{-0.70}$
 & $3.95^{+0.85}_{-1.02}$
 & $3.37 \pm 0.10$
\\
 & PyROA 
 & $--$
 & $-0.00 \pm 0.04$
 & $1.23^{+0.48}_{-0.51}$
 & $3.17^{+0.51}_{-0.54}$
 & $4.00^{+0.63}_{-0.62}$
 & $3.40 \pm 0.17$
\\
\addlinespace
Mrk 1044 & PyCCF 
 & $-0.39^{+1.34}_{-1.67}$
 & $0.00^{+0.64}_{-0.58}$
 & $-0.77^{+1.21}_{-1.72}$
 & $-0.46^{+1.28}_{-3.34}$
 & $0.06^{+3.93}_{-7.57}$
 & $-0.54 \pm 0.49$
\\
 & PyROA 
 & $-0.61^{+1.29}_{-0.94}$
 & $0.12^{+1.40}_{-1.18}$
 & $1.48^{+1.96}_{-1.78}$
 & $1.65^{+1.92}_{-1.67}$
 & $3.08^{+2.06}_{-1.81}$
 & $2.34 \pm 0.21$
\\
\addlinespace
\enddata
\tablecomments{All measurements are given in units of days. Object Lags are displayed in rest-frame.}
\centering
\end{deluxetable*}

\begin{deluxetable*}{lccccccc}
\tablecolumns{8}
\tablecaption{2019-2020 Time Lags
\label{table:year2_lags}}
\tablehead{
Object & Method & $u$ & $g$ & $r$ & $i$ & $z$ & $\tau_0$ }
\startdata
3C390.3 & PyCCF 
 & $7.10^{+14.17}_{-4.29}$
 & $0.03^{+3.69}_{-3.74}$
 & $0.44^{+4.82}_{-4.26}$
 & $6.96^{+4.26}_{-5.13}$
 & $4.04^{+12.50}_{-6.81}$
 & $5.16 \pm 2.47$
\\
 & PyROA 
 & $3.60^{+5.75}_{-4.92}$
 & $0.04^{+0.48}_{-0.51}$
 & $-1.53^{+0.75}_{-0.79}$
 & $5.89^{+0.95}_{-1.08}$
 & $6.24^{+1.57}_{-1.61}$
 & $4.22 \pm 1.99$
\\
\addlinespace
Arp 151 & PyCCF 
 & $-8.70^{+2.74}_{-4.62}$
 & $0.00^{+0.60}_{-0.58}$
 & $4.96^{+8.51}_{-6.01}$
 & $2.69^{+11.32}_{-2.62}$
 & $2.62^{+4.09}_{-9.61}$
 & $6.29 \pm 4.17$
\\
 & PyROA 
 & $-6.81^{+4.35}_{-3.48}$
 & $-0.00 \pm 0.03$
 & $1.64^{+12.69}_{-3.96}$
 & $0.24^{+3.36}_{-3.49}$
 & $0.89^{+4.00}_{-4.60}$
 & $1.53 \pm 2.13$
\\
\addlinespace
Mrk 6 & PyCCF 
 & $5.62^{+3.29}_{-3.19}$
 & $0.03^{+2.12}_{-2.18}$
 & $2.30^{+2.43}_{-2.28}$
 & $1.57^{+2.69}_{-2.43}$
 & $-0.05^{+2.45}_{-2.84}$
 & $0.53 \pm 1.70$
\\
 & PyROA 
 & $2.37^{+1.27}_{-1.28}$
 & $0.05 \pm 0.54$
 & $2.66^{+0.66}_{-0.64}$
 & $0.48 \pm 0.72$
 & $2.20 \pm 0.69$
 & $1.72 \pm 0.89$
\\
\addlinespace
Mrk 335 & PyCCF 
 & $-0.64^{+0.85}_{-0.87}$
 & $-0.00^{+0.85}_{-0.84}$
 & $0.60^{+0.88}_{-0.82}$
 & $0.98^{+0.73}_{-0.86}$
 & $1.40^{+0.96}_{-0.85}$
 & $1.20 \pm 0.09$
\\
 & PyROA 
 & $-0.23^{+0.21}_{-0.22}$
 & $-0.00 \pm 0.11$
 & $0.82 \pm 0.21$
 & $1.27 \pm 0.26$
 & $1.40^{+0.55}_{-0.54}$
 & $1.42 \pm 0.18$
\\
\addlinespace
Mrk 876 & PyCCF 
 & $-1.77^{+3.20}_{-3.97}$
 & $0.01^{+1.44}_{-1.45}$
 & $1.47^{+2.08}_{-1.63}$
 & $15.12^{+2.74}_{-3.88}$
 & $7.64^{+4.49}_{-3.54}$
 & $9.01 \pm 2.95$
\\
 & PyROA 
 & $-2.09^{+1.21}_{-1.22}$
 & $0.01^{+0.12}_{-0.11}$
 & $0.79^{+0.38}_{-0.36}$
 & $4.68^{+0.69}_{-0.65}$
 & $9.11^{+0.90}_{-0.93}$
 & $5.25 \pm 1.11$
\\
\addlinespace
NGC 3516 & PyCCF 
 & $-2.74^{+2.23}_{-1.97}$
 & $-0.01^{+2.84}_{-2.81}$
 & $6.84^{+2.34}_{-3.64}$
 & $0.41^{+3.27}_{-3.46}$
 & $7.51^{+3.13}_{-3.14}$
 & $5.74 \pm 1.98$
\\
 & PyROA 
 & $-1.23 \pm 0.33$
 & $0.00 \pm 0.02$
 & $0.45^{+0.39}_{-0.37}$
 & $0.49 \pm 0.42$
 & $1.09^{+0.46}_{-0.47}$
 & $1.03 \pm 0.40$
\\
\addlinespace
PG0052+251 & PyCCF 
 & $2.92^{+3.94}_{-3.58}$
 & $-0.02^{+1.63}_{-1.60}$
 & $9.47^{+4.02}_{-6.16}$
 & $8.42^{+3.77}_{-6.40}$
 & $13.16^{+4.18}_{-4.82}$
 & $9.95 \pm 2.86$
\\
 & PyROA 
 & $1.94^{+0.91}_{-0.92}$
 & $0.01 \pm 0.09$
 & $3.60^{+0.78}_{-0.83}$
 & $2.60^{+0.84}_{-0.97}$
 & $7.82^{+1.20}_{-1.25}$
 & $4.67 \pm 1.61$
\\
\addlinespace
PG0804+761 & PyCCF 
 & $-0.77^{+3.05}_{-3.37}$
 & $-0.03^{+2.25}_{-2.29}$
 & $6.88^{+5.55}_{-4.02}$
 & $19.13^{+8.86}_{-7.43}$
 & $33.14^{+4.41}_{-5.44}$
 & $22.74 \pm 3.98$
\\
 & PyROA 
 & $0.30^{+0.99}_{-0.92}$
 & $-0.00 \pm 0.03$
 & $1.57^{+0.79}_{-0.76}$
 & $13.34^{+2.38}_{-4.21}$
 & $19.21^{+1.89}_{-1.58}$
 & $10.07 \pm 3.34$
\\
\addlinespace
\enddata
\tablecomments{All measurements are given in units of days. Object Lags are displayed in rest-frame.}
\centering
\end{deluxetable*}

\begin{deluxetable*}{lccccccc}
\tablecolumns{8}
\tablecaption{2020-2021 Time Lags
\label{table:year3_lags}}
\tablehead{
Object & Method & $u$ & $g$ & $r$ & $i$ & $z$ & $\tau_0$ }
\startdata
3C390.3 & PyCCF 
 & $39.12^{+4.73}_{-32.38}$
 & $0.00^{+3.50}_{-3.48}$
 & $3.65^{+3.23}_{-3.16}$
 & $4.32^{+3.65}_{-3.33}$
 & $42.85^{+5.01}_{-7.07}$
 & $15.79 \pm 7.76$
\\
 & PyROA 
 & $46.53^{+0.71}_{-41.89}$
 & $-0.01^{+0.27}_{-0.28}$
 & $0.13^{+0.86}_{-0.80}$
 & $0.64^{+0.74}_{-0.81}$
 & $7.79^{+2.51}_{-2.24}$
 & $1.55 \pm 1.34$
\\
\addlinespace
MCG+8-11-11 & PyCCF 
 & $0.21^{+0.63}_{-0.61}$
 & $0.00^{+0.36}_{-0.37}$
 & $2.35^{+0.44}_{-0.41}$
 & $2.54^{+0.39}_{-0.40}$
 & $3.02^{+0.52}_{-0.50}$
 & $2.86 \pm 0.50$
\\
 & PyROA 
 & $0.17 \pm 0.16$
 & $-0.00 \pm 0.06$
 & $1.42 \pm 0.09$
 & $1.73 \pm 0.09$
 & $1.95^{+0.15}_{-0.16}$
 & $1.98 \pm 0.33$
\\
\addlinespace
Mrk 6 & PyCCF 
 & $-1.91^{+3.14}_{-2.74}$
 & $-0.02^{+1.28}_{-1.23}$
 & $2.89^{+1.21}_{-1.16}$
 & $0.28 \pm 1.13$
 & $1.64^{+1.30}_{-1.26}$
 & $1.45 \pm 0.85$
\\
 & PyROA 
 & $-1.96^{+0.60}_{-0.63}$
 & $-0.00 \pm 0.16$
 & $0.99^{+0.19}_{-0.17}$
 & $0.13^{+0.17}_{-0.19}$
 & $0.28^{+0.23}_{-0.22}$
 & $0.43 \pm 0.37$
\\
\addlinespace
Mrk 50 & PyCCF 
 & $-0.34^{+0.77}_{-0.69}$
 & $-0.00 \pm 0.38$
 & $1.30^{+0.45}_{-0.46}$
 & $1.28 \pm 0.56$
 & $1.57^{+0.80}_{-0.72}$
 & $1.62 \pm 0.31$
\\
 & PyROA 
 & $-0.25^{+0.18}_{-0.16}$
 & $-0.00 \pm 0.03$
 & $0.45 \pm 0.10$
 & $0.63 \pm 0.13$
 & $1.03 \pm 0.21$
 & $0.83 \pm 0.06$
\\
\addlinespace
Mrk 279 & PyCCF 
 & $-0.14^{+0.64}_{-0.62}$
 & $-0.01 \pm 0.44$
 & $1.57^{+0.54}_{-0.53}$
 & $0.91^{+0.61}_{-0.58}$
 & $3.13^{+0.89}_{-0.74}$
 & $2.01 \pm 0.49$
\\
 & PyROA 
 & $-0.21 \pm 0.30$
 & $0.00 \pm 0.07$
 & $1.28^{+0.22}_{-0.20}$
 & $1.12 \pm 0.22$
 & $2.58^{+0.35}_{-0.35}$
 & $1.81 \pm 0.30$
\\
\addlinespace
Mrk 817 & PyCCF 
 & $-0.89^{+1.34}_{-1.26}$
 & $0.00 \pm 0.53$
 & $1.89^{+0.85}_{-0.82}$
 & $3.32^{+0.76}_{-0.82}$
 & $4.18^{+0.99}_{-1.08}$
 & $3.70 \pm 0.19$
\\
 & PyROA 
 & $-0.47^{+0.27}_{-0.26}$
 & $-0.00 \pm 0.04$
 & $1.96 \pm 0.26$
 & $2.32 \pm 0.25$
 & $3.66 \pm 0.31$
 & $2.96 \pm 0.30$
\\
\addlinespace
Mrk 1501 & PyCCF 
 & $--$
 & $0.02^{+1.99}_{-1.97}$
 & $1.40^{+2.20}_{-2.23}$
 & $7.72^{+2.62}_{-3.31}$
 & $10.40^{+3.03}_{-9.77}$
 & $7.61 \pm 1.39$
\\
 & PyROA 
 & $--$
 & $0.02^{+0.33}_{-0.32}$
 & $1.56^{+1.05}_{-1.07}$
 & $16.19^{+1.07}_{-1.20}$
 & $8.45^{+2.06}_{-1.76}$
 & $12.47 \pm 3.86$
\\
\addlinespace
NGC 7469 & PyCCF 
 & $0.64 \pm 1.10$
 & $0.01^{+0.97}_{-0.96}$
 & $2.84^{+1.12}_{-1.14}$
 & $3.47^{+1.33}_{-1.19}$
 & $5.68^{+1.68}_{-1.32}$
 & $4.23 \pm 0.90$
\\
 & PyROA 
 & $0.12^{+0.27}_{-0.29}$
 & $-0.00 \pm 0.03$
 & $1.05 \pm 0.27$
 & $1.19 \pm 0.29$
 & $2.43^{+0.44}_{-0.45}$
 & $1.64 \pm 0.32$
\\
\addlinespace
PG2130+099 & PyCCF 
 & $-0.54^{+1.40}_{-1.69}$
 & $-0.00^{+0.68}_{-0.67}$
 & $2.19^{+0.66}_{-0.69}$
 & $7.69^{+1.37}_{-2.09}$
 & $4.70^{+1.83}_{-1.93}$
 & $5.36 \pm 1.06$
\\
 & PyROA 
 & $-0.90^{+0.73}_{-0.72}$
 & $-0.00 \pm 0.15$
 & $1.70^{+0.31}_{-0.30}$
 & $15.99^{+1.28}_{-9.24}$
 & $6.45 \pm 0.92$
 & $4.60 \pm 0.69$
\\
\addlinespace
\enddata
\tablecomments{All measurements are given in units of days. Object Lags are displayed in rest-frame.}
\centering
\end{deluxetable*}

\begin{deluxetable*}{lccccccc}
\tablecolumns{8}
\tablecaption{2021-2022 Time Lags
\label{table:year4_lags}}
\tablehead{
Object & Method & $u$ & $g$ & $r$ & $i$ & $z$ & $\tau_0$ }
\startdata
MCG+8-11-11 & PyCCF 
 & $-0.21^{+0.72}_{-0.67}$
 & $-0.00^{+0.55}_{-0.56}$
 & $2.95^{+0.64}_{-0.59}$
 & $3.12^{+0.59}_{-0.53}$
 & $4.64 \pm 0.72$
 & $3.88 \pm 0.55$
\\
 & PyROA 
 & $-0.14^{+0.22}_{-0.23}$
 & $0.00^{+0.06}_{-0.07}$
 & $1.69^{+0.16}_{-0.15}$
 & $2.12 \pm 0.16$
 & $3.18 \pm 0.18$
 & $2.63 \pm 0.27$
\\
\addlinespace
Mrk 110 & PyCCF 
 & $0.67^{+0.54}_{-0.53}$
 & $-0.00^{+0.33}_{-0.32}$
 & $1.65 \pm 0.40$
 & $1.45 \pm 0.39$
 & $2.26^{+0.70}_{-0.71}$
 & $1.87 \pm 0.53$
\\
 & PyROA 
 & $0.91 \pm 0.30$
 & $0.00 \pm 0.13$
 & $1.75 \pm 0.23$
 & $1.55 \pm 0.23$
 & $2.59^{+0.40}_{-0.38}$
 & $2.02 \pm 0.61$
\\
\addlinespace
Mrk 142 & PyCCF 
 & $-1.39^{+1.43}_{-2.53}$
 & $0.00 \pm 0.71$
 & $1.79^{+2.56}_{-1.35}$
 & $3.72^{+1.71}_{-2.10}$
 & $14.52^{+9.79}_{-8.27}$
 & $4.84 \pm 0.91$
\\
 & PyROA 
 & $-0.47^{+0.76}_{-0.82}$
 & $-0.00 \pm 0.06$
 & $1.11^{+0.35}_{-0.32}$
 & $2.39^{+0.44}_{-0.40}$
 & $8.05^{+5.86}_{-4.40}$
 & $2.78 \pm 0.23$
\\
\addlinespace
Mrk 279 & PyCCF 
 & $1.92^{+1.01}_{-0.96}$
 & $-0.02^{+0.52}_{-0.50}$
 & $-0.71^{+0.84}_{-0.82}$
 & $0.13^{+0.83}_{-0.72}$
 & $5.13^{+1.05}_{-0.96}$
 & $1.76 \pm 1.32$
\\
 & PyROA 
 & $1.51 \pm 0.37$
 & $0.00^{+0.15}_{-0.16}$
 & $-0.51^{+0.41}_{-0.37}$
 & $0.29^{+0.34}_{-0.33}$
 & $4.21^{+0.43}_{-0.40}$
 & $1.60 \pm 1.07$
\\
\addlinespace
Mrk 335 & PyCCF 
 & $5.89^{+2.25}_{-2.30}$
 & $0.01 \pm 2.32$
 & $-0.12^{+3.48}_{-3.66}$
 & $0.28^{+2.33}_{-2.57}$
 & $-0.06^{+3.52}_{-3.86}$
 & $-1.30 \pm 2.43$
\\
 & PyROA 
 & $3.80^{+0.94}_{-0.91}$
 & $0.01 \pm 0.16$
 & $-0.80^{+0.75}_{-0.74}$
 & $0.88^{+0.85}_{-0.82}$
 & $-0.06^{+1.05}_{-1.01}$
 & $-0.37 \pm 1.27$
\\
\addlinespace
Mrk 50 & PyCCF 
 & $0.64 \pm 0.97$
 & $-0.01^{+1.06}_{-1.04}$
 & $2.39^{+0.82}_{-1.07}$
 & $1.23^{+1.01}_{-1.31}$
 & $1.22^{+1.80}_{-1.66}$
 & $1.62 \pm 0.96$
\\
 & PyROA 
 & $0.88^{+0.41}_{-0.44}$
 & $0.00 \pm 0.09$
 & $0.88^{+0.27}_{-0.25}$
 & $0.44^{+0.29}_{-0.26}$
 & $0.98^{+0.79}_{-0.82}$
 & $0.71 \pm 0.45$
\\
\addlinespace
Mrk 817 & PyCCF 
 & $0.82^{+0.93}_{-0.88}$
 & $-0.01 \pm 0.57$
 & $1.98^{+0.64}_{-0.60}$
 & $2.87 \pm 0.69$
 & $2.40^{+1.22}_{-1.30}$
 & $2.89 \pm 0.73$
\\
 & PyROA 
 & $0.90^{+0.36}_{-0.37}$
 & $0.01^{+0.19}_{-0.20}$
 & $1.30^{+0.31}_{-0.32}$
 & $1.73^{+0.32}_{-0.29}$
 & $2.30 \pm 0.46$
 & $1.86 \pm 0.53$
\\
\addlinespace
PG2130+099 & PyCCF 
 & $-5.75^{+1.92}_{-1.78}$
 & $0.01^{+0.81}_{-0.82}$
 & $-1.13^{+0.93}_{-1.04}$
 & $7.84^{+1.68}_{-1.99}$
 & $2.48^{+1.54}_{-1.37}$
 & $3.04 \pm 2.18$
\\
 & PyROA 
 & $-4.29^{+0.74}_{-0.77}$
 & $-0.00 \pm 0.10$
 & $0.23^{+0.35}_{-0.37}$
 & $3.82 \pm 0.56$
 & $2.80^{+0.63}_{-0.61}$
 & $2.88 \pm 1.06$
\\
\addlinespace
\enddata
\tablecomments{All measurements are given in units of days. Object Lags are displayed in rest-frame.}
\centering
\end{deluxetable*}

\begin{deluxetable*}{cccccc}
\tablecolumns{6}
\tablecaption{Predicted and Measured $5100\Angstrom$ Continuum Emitting Sizes}
\label{table:disksizes}
\tablehead{
Object & Observation Year & $\tau_{0}$, $X=2.49$ & $\tau_{0}$, $X=5.04$ & $\tau_{5100}$ & Reduced $\chi^2$}
\startdata
3C390.3 &\makecell{$2019-2020$ \\$2020-2021$ } & \makecell{1.49\\1.56} & \makecell{3.82\\3.98} &\makecell{6.07 $\pm$~2.90\\18.56 $\pm$~9.12} &\makecell{0.35\\7.07} \\
\hline
Arp 151 &\makecell{$2019-2020$ } & \makecell{0.11} & \makecell{0.29} &\makecell{7.07 $\pm$~4.69} &\makecell{1.01} \\
\hline
MCG+8-11-11 &\makecell{$2020-2021$ \\$2021-2022$ } & \makecell{0.54\\0.56} & \makecell{1.38\\1.43} &\makecell{3.22 $\pm$~0.56\\4.36 $\pm$~0.61} &\makecell{2.91\\1.75} \\
\hline
Mrk 6 &\makecell{$2019-2020$ \\$2020-2021$ } & \makecell{0.78\\0.78} & \makecell{2.00\\1.99} &\makecell{0.59 $\pm$~1.91\\1.63 $\pm$~0.95} &\makecell{1.06\\1.16} \\
\hline
Mrk 50 &\makecell{$2020-2021$ \\$2021-2022$ } & \makecell{0.42\\0.40} & \makecell{1.07\\1.02} &\makecell{1.83 $\pm$~0.34\\1.83 $\pm$~1.08} &\makecell{0.55\\1.23} \\
\hline
Mrk 110 &\makecell{$2018-2019$ \\$2021-2022$ } & \makecell{0.53\\0.46} & \makecell{1.35\\1.18} &\makecell{4.19 $\pm$~0.70\\2.14 $\pm$~0.61} &\makecell{0.13\\2.62} \\
\hline
Mrk 142 &\makecell{$2018-2019$ \\$2021-2022$ } & \makecell{0.18\\0.16} & \makecell{0.45\\0.42} &\makecell{0.39 $\pm$~0.38\\5.61 $\pm$~1.06} &\makecell{0.19\\0.24} \\
\hline
Mrk 279 &\makecell{$2020-2021$ \\$2021-2022$ } & \makecell{0.33\\0.41} & \makecell{0.86\\1.04} &\makecell{2.29 $\pm$~0.56\\2.00 $\pm$~1.50} &\makecell{1.24\\5.34} \\
\hline
Mrk 335 &\makecell{$2019-2020$ \\$2021-2022$ } & \makecell{0.42\\0.40} & \makecell{1.08\\1.02} &\makecell{1.36 $\pm$~0.10\\-1.47 $\pm$~2.75} &\makecell{0.03\\1.57} \\
\hline
Mrk 817 &\makecell{$2018-2019$ \\$2020-2021$ \\$2021-2022$ } & \makecell{0.73\\0.76\\0.75} & \makecell{1.87\\1.96\\1.93} &\makecell{3.84 $\pm$~0.11\\4.21 $\pm$~0.21\\3.29 $\pm$~0.84} &\makecell{0.03\\0.10\\1.63} \\
\hline
Mrk 876 &\makecell{$2019-2020$ } & \makecell{2.09} & \makecell{5.36} &\makecell{11.57 $\pm$~3.79} &\makecell{1.90} \\
\hline
Mrk 1044 &\makecell{$2018-2019$ } & \makecell{0.17} & \makecell{0.43} &\makecell{-0.60 $\pm$~0.55} &\makecell{0.07} \\
\hline
Mrk 1501 &\makecell{$2020-2021$ } & \makecell{1.16} & \makecell{2.96} &\makecell{9.32 $\pm$~1.71} &\makecell{0.30} \\
\hline
NGC 3516 &\makecell{$2019-2020$ } & \makecell{0.20} & \makecell{0.52} &\makecell{6.35 $\pm$~2.19} &\makecell{1.03} \\
\hline
NGC 7469 &\makecell{$2020-2021$ } & \makecell{0.33} & \makecell{0.84} &\makecell{4.73 $\pm$~1.01} &\makecell{1.10} \\
\hline
PG0052+251 &\makecell{$2019-2020$ } & \makecell{2.01} & \makecell{5.16} &\makecell{13.18 $\pm$~3.79} &\makecell{0.95} \\
\hline
PG0804+761 &\makecell{$2019-2020$ } & \makecell{3.73} & \makecell{9.55} &\makecell{28.23 $\pm$~4.95} &\makecell{1.45} \\
\hline
PG2130+099 &\makecell{$2020-2021$ \\$2021-2022$ } & \makecell{0.91\\0.89} & \makecell{2.32\\2.27} &\makecell{6.36 $\pm$~1.26\\3.60 $\pm$~2.59} &\makecell{1.25\\5.43} \\
\enddata
\tablecomments{{Predicted accretion disk sizes of Eq.~\ref{eq:alpha} using different values of $X$ at 5100$\Angstrom$. The sizes we measure from observations are in the rightmost column and calculated using Eq.~\ref{eq:wavlag} with $\beta$ fixed at $4/3$, $\tau_0$ from Tables \ref{table:year1_lags} - \ref{table:year4_lags} respectively, $\lambda$ set to $5100\Angstrom$, and without the subtraction component of Eq.~\ref{eq:wavlag} such that the lag is not fixed to start at 0 from $\lambda_0$. All values are in units of days. These $\tau_{5100}$ values are plotted in Fig.~\ref{fig:lum_plot_both} except for those that are \textbf{consistent with zero or are used in the ICBM sample}.}}
\centering
\end{deluxetable*}

\movetabledown=6.5cm
\begin{rotatetable}
\begin{deluxetable*}{lcccccccc}
\tablecaption{Zowada Observation Target Information
\label{table:AGN_facts}}
\tablehead{
Object & Redshift & Distance (Mpc) & E(B-V) & Mass ($M_{\odot}$) & Observed log $L_{5100}$ (erg/s) & Host Galaxy log $L_{5100}$ (erg/s) & AGN log $L_{5100}$ (erg/s)}
\startdata
3C390.3 & 0.056 & 243.5 & 0.061 & 8.638 & \makecell{1.02e+44 $\pm$ 4.37e+42 ($2019-2020$)\\1.10e+44 $\pm$ 2.36e+43 ($2020-2021$)} & 3.81e+43 $\pm$ 1.91e+42 & \makecell{6.17e+43 $\pm$ 4.81e+42 ($2019-2020$)\\6.99e+43 $\pm$ 2.37e+43 ($2020-2021$)}\\
\hline
Arp 151 & 0.021 & 89.2 & 0.012 & 6.672 & \makecell{1.07e+43 $\pm$ 7.40e+41 ($2019-2020$)} & 7.93e+42 $\pm$ 3.97e+41 & \makecell{2.57e+42 $\pm$ 8.43e+41 ($2019-2020$)}\\
\hline
MCG+8-11-11 & 0.020 & 86.6 & 0.184 & 7.288 & \makecell{6.70e+43 $\pm$ 8.91e+42 ($2020-2021$)\\7.48e+43 $\pm$ 7.42e+42 ($2021-2022$)} & --- & ---\\
\hline
Mrk 6 & 0.019 & 80.6 & 0.117 & 8.102 & \makecell{3.12e+43 $\pm$ 4.46e+42 ($2019-2020$)\\3.07e+43 $\pm$ 4.39e+42 ($2020-2021$)} & --- & ---\\
\hline
Mrk 50 & 0.023 & 100.7 & 0.014 & 7.422 & \makecell{1.73e+43 $\pm$ 1.79e+42 ($2020-2021$)\\1.49e+43 $\pm$ 1.41e+42 ($2021-2022$)} & --- & ---\\
\hline
Mrk 110 & 0.035 & 150.9 & 0.011 & 7.292 & \makecell{7.73e+43 $\pm$ 5.32e+42 ($2018-2019$)\\5.47e+43 $\pm$ 6.61e+42 ($2021-2022$)} & 1.07e+43 $\pm$ 5.36e+41 & \makecell{6.62e+43 $\pm$ 5.35e+42 ($2018-2019$)\\4.36e+43 $\pm$ 6.63e+42 ($2021-2022$)}\\
\hline
Mrk 142 & 0.045 & 193.5 & 0.014 & 6.294 & \makecell{4.56e+43 $\pm$ 2.45e+42 ($2018-2019$)\\4.05e+43 $\pm$ 3.69e+42 ($2021-2022$)} & 1.91e+43 $\pm$ 9.56e+41 & \makecell{2.57e+43 $\pm$ 2.65e+42 ($2018-2019$)\\2.05e+43 $\pm$ 3.82e+42 ($2021-2022$)}\\
\hline
Mrk 279 & 0.030 & 129.7 & 0.014 & 7.435 & \makecell{4.86e+43 $\pm$ 3.97e+42 ($2020-2021$)\\5.81e+43 $\pm$ 6.07e+42 ($2021-2022$)} & 3.56e+43 $\pm$ 1.78e+42 & \makecell{1.19e+43 $\pm$ 4.38e+42 ($2020-2021$)\\2.14e+43 $\pm$ 6.34e+42 ($2021-2022$)}\\
\hline
Mrk 335 & 0.026 & 109.5 & 0.030 & 7.230 & \makecell{5.21e+43 $\pm$ 4.22e+42 ($2019-2020$)\\4.59e+43 $\pm$ 2.64e+42 ($2021-2022$)} & 1.43e+43 $\pm$ 7.14e+41 & \makecell{3.74e+43 $\pm$ 4.29e+42 ($2019-2020$)\\3.12e+43 $\pm$ 2.74e+42 ($2021-2022$)}\\
\hline
Mrk 817 & 0.031 & 134.2 & 0.022 & 7.586 & \makecell{1.11e+44 $\pm$ 8.42e+42 ($2018-2019$)\\1.24e+44 $\pm$ 1.12e+43 ($2020-2021$)\\1.19e+44 $\pm$ 1.29e+43 ($2021-2022$)} & 2.24e+43 $\pm$ 1.12e+42 & \makecell{8.83e+43 $\pm$ 8.50e+42 ($2018-2019$)\\1.01e+44 $\pm$ 1.13e+43 ($2020-2021$)\\9.59e+43 $\pm$ 1.30e+43 ($2021-2022$)}\\
\hline
Mrk 876 & 0.129 & 588.4 & 0.023 & 8.339 & \makecell{6.52e+44 $\pm$ 2.21e+43 ($2018-2019$)\\7.95e+44 $\pm$ 6.45e+43 ($2019-2020$)} & 2.41e+44 $\pm$ 1.20e+43 & \makecell{3.81e+44 $\pm$ 2.60e+43 ($2018-2019$)\\5.23e+44 $\pm$ 6.59e+43 ($2019-2020$)}\\
\hline
Mrk 1044 & 0.016 & 70.3 & 0.029 & 6.146 & \makecell{2.68e+43 $\pm$ 1.54e+42 ($2018-2019$)} & --- & ---\\
\hline
Mrk 1501 & 0.089 & 402.5 & 0.085 & 8.067 & \makecell{1.44e+44 $\pm$ 8.46e+42 ($2020-2021$)} & --- & ---\\
\hline
NGC 3516 & 0.009 & 37.1 & 0.036 & 7.399 & \makecell{1.83e+43 $\pm$ 1.66e+42 ($2019-2020$)} & 1.56e+43 $\pm$ 1.56e+42 & \makecell{2.59e+42 $\pm$ 2.28e+42 ($2019-2020$)}\\
\hline
NGC 7469 & 0.016 & 68.8 & 0.060 & 6.956 & \makecell{6.47e+43 $\pm$ 5.35e+42 ($2020-2021$)} & 3.22e+43 $\pm$ 3.22e+42 & \makecell{3.21e+43 $\pm$ 6.27e+42 ($2020-2021$)}\\
\hline
PG0052+251 & 0.154 & 715.9 & 0.040 & 8.462 & \makecell{5.98e+44 $\pm$ 6.77e+43 ($2019-2020$)} & 1.84e+44 $\pm$ 9.21e+42 & \makecell{3.85e+44 $\pm$ 6.85e+43 ($2019-2020$)}\\
\hline
PG0804+761 & 0.100 & 447.5 & 0.030 & 8.735 & \makecell{1.15e+45 $\pm$ 1.08e+44 ($2019-2020$)} & 6.67e+43 $\pm$ 3.33e+42 & \makecell{1.07e+45 $\pm$ 1.08e+44 ($2019-2020$)}\\
\hline
PG2130+099 & 0.063 & 274.7 & 0.037 & 7.433 & \makecell{3.03e+44 $\pm$ 1.36e+43 ($2020-2021$)\\2.85e+44 $\pm$ 1.88e+43 ($2021-2022$)} & 3.10e+43 $\pm$ 1.55e+42 & \makecell{2.70e+44 $\pm$ 1.37e+43 ($2020-2021$)\\2.52e+44 $\pm$ 1.88e+43 ($2021-2022$)}\\
\hline
\enddata
\tablecomments{Information for each AGN targetting by the Zowada. Masses, luminosity distances, and redshifts come from the AGN Black Hole Mass Database \citep{bentz2015}. The only exception is Mrk~1044's mass, which is taken from \cite{wang2014}.}
\centering
\end{deluxetable*}
\end{rotatetable}


\begin{figure*}[hbt!]
 \centering
 \includegraphics[trim=0.5cm 0cm 0.5cm 0.5cm, clip = true,width=0.99\textwidth]{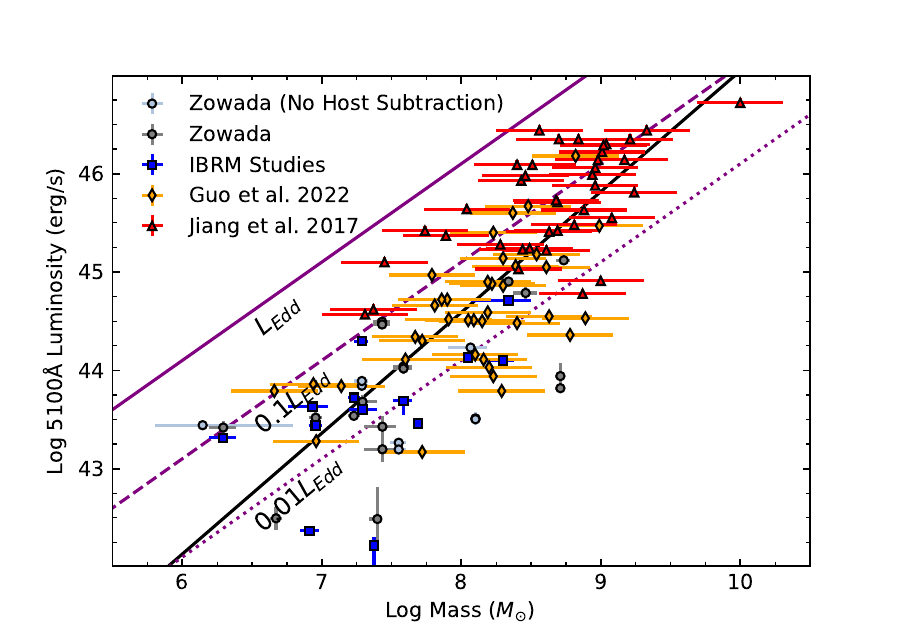}
 \caption{The relationship between log SMBH mass and log AGN luminosity. {Gray circles and light-gray circles} are Zowada objects with and without host subtraction, respectively. Blue squares are the {results from} IBRM studies, orange diamonds are from \cite{guo2022}'s ZTF study, and the red triangles are from \cite{2017ApJ...836..186J}'s Pan-STARRS objects. The black line is the best fitting linear relation to the log of both mass and luminosity. {We find a slope of $1.24\pm0.10$, slightly larger than that} found by \cite{guo2022}. {The purple solid, dashed, and dotted lines represents Eddington, 10\% Eddington, and 1\% Eddington luminosity ($L_{Edd}$), respectively.}}
 \label{fig:massvlum}
 \end{figure*}


\begin{figure*}[hbt!]
\figurenum{7}
\plotone{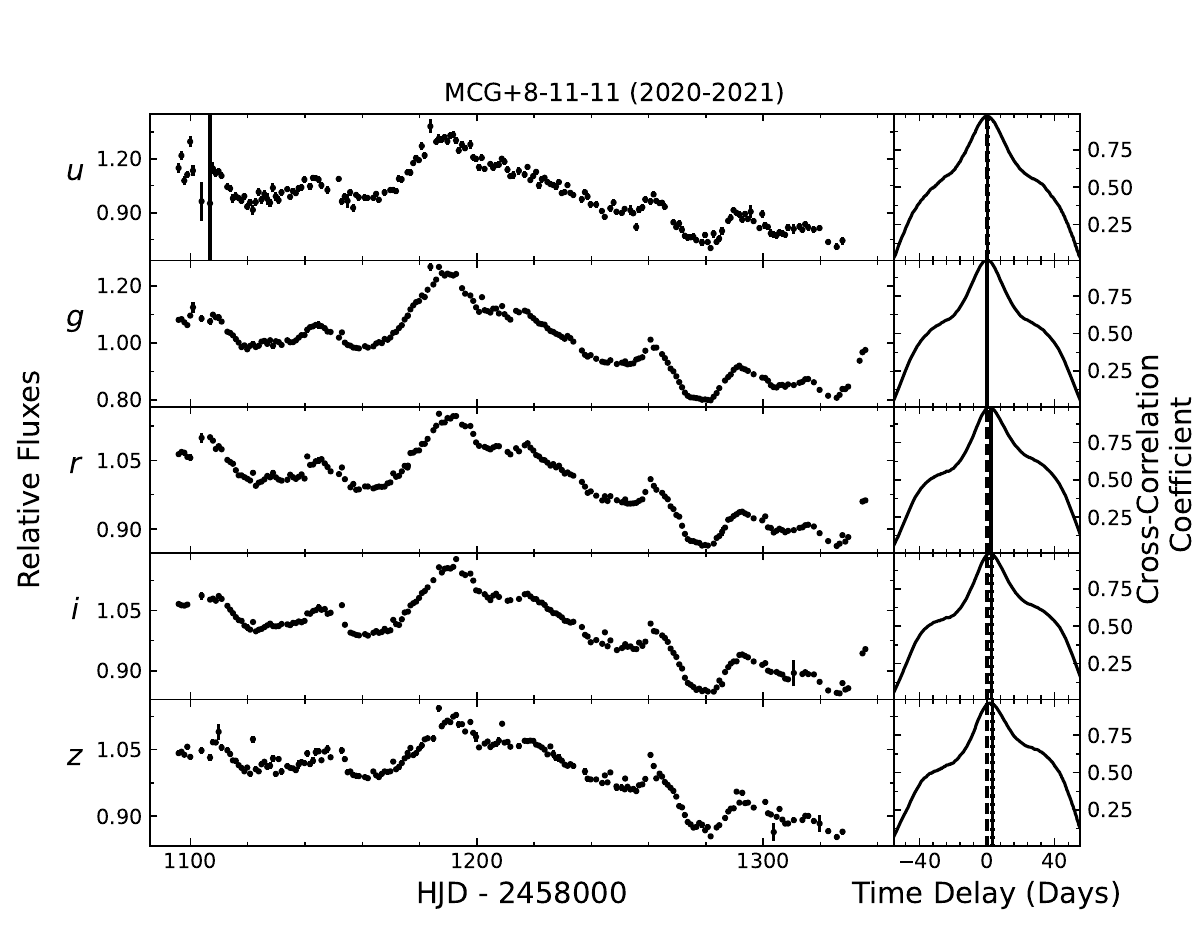}
\caption{The relative photometry light curves and measured CCFs for our entire sample. From top to bottom the $ugriz$ light curves are presented. The right panel is the CCF measured with respect to the $g$ band, with the solid black vertical line representing the lag measurement. The dotted black lines are the uncertainties, and the dashed black line is fixed to 0 for reference. The complete figure set (29~images) will be made available in the online journal once published.}
\end{figure*}

\clearpage
\bibliography{YearCampaign.bib}{}
\bibliographystyle{mnras}

\end{document}